\documentclass[twocolumn]{aastex63} 

\usepackage{amsmath}
\usepackage{amssymb}
\usepackage{empheq}
\usepackage{mathrsfs}
\usepackage{enumitem}
				
\usepackage{amssymb}

\shorttitle{Migration Pathways of Warm Jupiters}
\shortauthors{Morgan et al.}
\graphicspath{{./}{figures/}}
\begin{document}


\title{Exploring Warm Jupiter Migration Pathways with Eccentricities II. \\ 
Correlations with Host Star Properties and Orbital Separation}

\correspondingauthor{Marvin Morgan}
\email{marv08@utexas.edu }

\author[0000-0003-4022-6234]{Marvin Morgan}
\affiliation{Department of Astronomy, The University of Texas at Austin, Austin, TX 78712, USA}
\affiliation{Department of Physics, University of California, Santa Barbara, Santa Barbara, CA 93106, USA}

\author[0000-0003-2649-2288]{Brendan P. Bowler}
\affiliation{Department of Physics, University of California, Santa Barbara, Santa Barbara, CA 93106, USA}
\affiliation{Department of Astronomy, The University of Texas at Austin, Austin, TX 78712, USA}

\author[0000-0001-6532-6755]{Quang H. Tran}
\affiliation{Department of Astronomy, Yale University, New Haven, CT 06511, USA}
\affiliation{Department of Astronomy, The University of Texas at Austin, Austin, TX 78712, USA}

\begin{abstract}

Warm Jupiters with orbital periods of $\approx$10-365 d represent a population of giant planets located well within the water ice line but beyond the region of tidal influence of their host star relevant for high-eccentricity tidal migration. Orbital eccentricities offer important clues about the formation and dynamical history of warm Jupiters because \emph{in situ} formation and disk migration should imprint near-circular orbits whereas planet scattering should excite eccentricities. Based on uniform Keplerian fits of 18,587 RVs targeting 200 warm Jupiters, we use hierarchical Bayesian modeling to evaluate the impact of host star metallicity, stellar mass, and orbital separation on the reconstructed population-level eccentricity distributions. Warm Jupiters take on a broad range of eccentricities, and their population-level eccentricities are well modeled using a Beta distribution with $\alpha$ = 1.00$^{+0.09}_{-0.08}$ and $\beta$ = 2.79$^{+0.28}_{-0.26}$. We find that 27$^{+3}_{-4}\%$ of warm Jupiters have eccentricities consistent with near-circular orbits ($e$ $<$ 0.1), suggesting that most warm Jupiters (73$^{+3}_{-3}\%$) detected are dynamically hot. Warm Jupiters orbiting metal-rich stars are more eccentric than those orbiting metal-poor stars---in agreement with earlier findings---but no differences are observed as a function of stellar host mass or orbital separation, at least within the characteristic ranges probed by our sample ($\approx$0.5--2.0 $M_{\odot}$ and 0.1--1 AU, respectively). In this sense, metallicity plays a larger role in shaping the underlying eccentricity distribution of warm Jupiters than stellar mass or final orbital distance. These results are broadly consistent with planet scattering playing a major role in shaping the orbital architectures of close-in giant planets.

\end{abstract}
\section{Introduction} \label{sec:intro}

The discovery of 51 Pegasi b (\citealt{Mayor1995}), and other early radial velocity planet detections (e.g. \citealt{Campbell1988}; \citealt{Cochran1997}), revolutionized assumptions of how giant planets form and migrate because prior knowledge of planetary architectures was conditioned on patterns seen in the Solar System. Ground-based radial velocity (RV) surveys have since shown that giant planets are found with a wide range of orbital configurations, overhauling previous paradigms of planet formation and evolution (\citealt{Santos2003}; \citealt{Udry2003}; \citealt{Butler2006}). Giant planets populate a wide range of separations spanning a few stellar radii to thousands of AU, with eccentricities ranging from circular to 0.95  (\citealt{Jones2006}; \citealt{Naef2010}; \citealt{Naud2014}; \citealt{WinnFabrycky2015}; \citealt{DawsonJohnson2018}; \citealt{Zhang2021}; \citealt{Gupta2024}). This diversity of orbital properties reflects a wide variety of physical and dynamical processes related to the conditions of their formation and their subsequent gravitational interactions.

Giant planets are expected to predominantly form where conditions are favorable for the rapid assembly of cores (\citealt{Pollack1996}). For Sun-like stars, the condensation of water into solid ice between $\sim$2–3 AU facilitates this process of core formation and subsequent gas accretion (\citealt{Lecar2006}; \citealt{KennedyKenyon2008}). This general picture is supported by an abrupt rise in the occurrence rate of giant planets which peaks at orbital distances of 1--10 AU (\citealt{Cumming2008}; \citealt{Fernandes2019}; \citealt{Wittenmyer2020}; \citealt{Fulton2021}) and falls at wider separations (\citealt{Bowler2016}; \citealt{Nielsen2019}).

The closest region within 1 AU offers an especially rich probe of planet formation and dynamical evolution. This region is traditionally separated into two populations of short-period giant planets based on their potential to tidally interact with their host stars. Historically, more attention has been paid to hot Jupiters (with orbital periods, $P$, conventionally restricted to $<$10 days), in part because of their relative ease of detection and the ability to constrain stellar obliquities for transiting systems (\citealt{Howard2010}; \citealt{Winn2010}; \citealt{Albrecht2012}; \citealt{Wright2012}). However, the origin of warm Jupiters (orbital periods of $\approx$10--365 days) has also proven to be an especially challenging question to address. Considerable effort has been dedicated to understanding the migration channels that transport these planets to their current orbital locations between $\approx$0.1--1 AU. Giant planets that form early enough can migrate from beyond the ice line through interactions with the gaseous protoplanetary disk (\citealt{GoldreichTremaine1980};  \citealt{Lin1996}; \citealt{KleyNelson2012}), or interactions with a planetesimal disk (\citealt{Murray1998}). High eccentricity migration through planet-planet scattering (\citealt{RasioFord1996}; \citealt{IdaLin2004}; \citealt{Chatterjee2008}; \citealt{BeaugNesvorn2012}; \citealt{Petrovich2015}; \citealt{DawsonJohnson2018}), von Zeipel-Lidov-Kozai (ZLK) oscillations with an outer companion (\citealt{Kozai1962}; \citealt{Lidov1962}; \citealt{WuMurray2003}; \citealt{Naoz2016}; \citealt{ItoOhtsuka2019}), and secular chaos (\citealt{WuLithwick2011}) may also be responsible for the observed population of warm Jupiters.
Alternatively, warm Jupiters may have avoided long-range orbital migration completely and formed \emph{in situ} under favorable conditions (\citealt{Batygin2016}; \citealt{Boley2016}). 

Eccentricities offer a potential way to distinguish the relative importance of these transport mechanisms. Disk migration and \emph{in situ} formation should retain low eccentricities (\citealt{Dunhill2013}; \citealt{Baruteau2014})
 while scattering events are expected to produce a broad range of eccentricities (\citealt{JuricTremaine2008}; \citealt{PetrovichTremaine2016}). ZLK oscillations can periodically excite eccentricities and shrink periastron distances, which eventually drives a planet close enough to its host star to activate tides. Tidal friction can then dissipate orbital energy and circularize the planet's orbit (\citealt{Eggleton2001}; \citealt{FabryckyTremaine2007}; \citealt{Wu2007}). Because this process is not instantaneous, highly eccentric warm Jupiters that migrate through constant angular momentum migration tracks should be observable (\citealt{Socrates2012}). However, this  pathway has been ruled out as a dominant migration mechanism for most warm Jupiters due to a relative dearth of observed highly-eccentric proto-hot Jupiters (\citealt{Dawson2015}; \citealt{Jackson2023}). In addition, warm Jupiters beyond $\approx$0.1 AU are too far from their host star to raise dissipative tides as part of ZLK-induced high-eccentricity migration.

In principle, then, measuring the eccentricities of warm Jupiters should give insight into the origin of this population. However, the interpretation of this eccentricity distribution is made difficult because warm Jupiters often have eccentricities too high for \emph{in situ} formation and disk migration, and too low for planet-planet scattering alone (\citealt{DawsonJohnson2018}). One way to disentangle the relative importance of these mechanisms is to examine how eccentricities of warm Jupiters may change with properties of the host star or orbital properties of the companion. For instance, if planet scattering is common and more massive stars produce more giant planets, these might be expected to dynamically interact more regularly and excite eccentricities more frequently compared to lower mass stars. In this way, uncovering how host star properties can affect the eccentricities of giant planets can provide direct constraints on migration and formation processes.

Previous studies examining the overall eccentricity distribution of giant planets have generally made use of heterogeneous compilations of individual eccentricity constraints based on a variety of approaches to fit RVs, and also assumed Gaussian uncertainties or no uncertainties at all (e.g. \citealt{Kipping2013}; \citealt{KaneWittenmyer2024}). In this work, we present results of the largest self-consistent and most statistically rigorous analysis of warm Jupiter eccentricities to date. Using hierarchical Bayesian modeling (HBM), we explore how the underlying eccentricity distribution of warm Jupiters varies with host star mass, metallicity, and orbital distance to understand how, when, and under what conditions giant planets undergo inward orbital migration.

\begin{figure*}
\begin{center}
{\includegraphics[width=\linewidth]{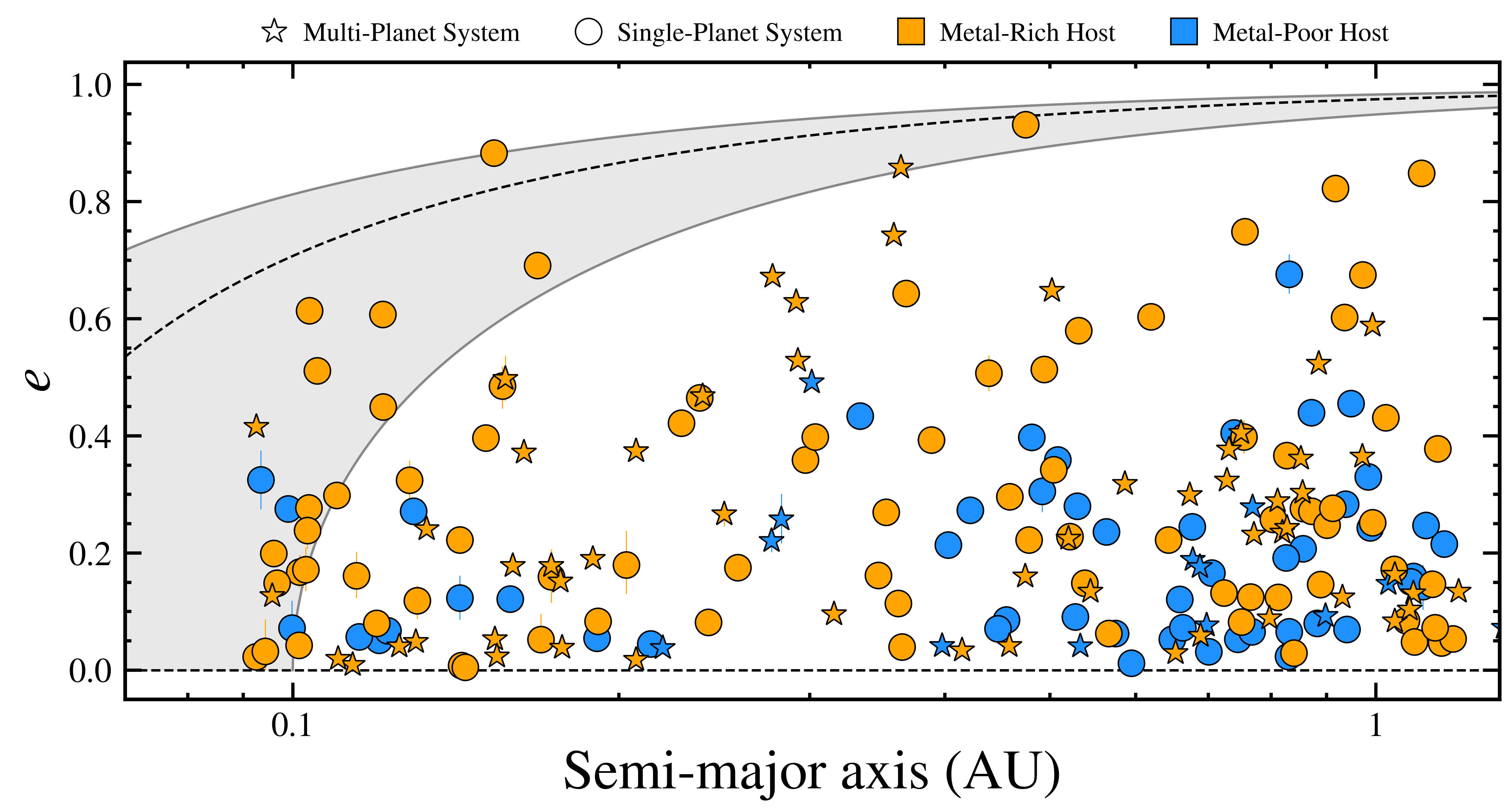}}
\caption{ Eccentricity plotted as a function of semi-major axis for all 200 warm Jupiters with new, consistently fit orbits in this analysis. Circles correspond to warm Jupiters in single planet systems and stars represent warm Jupiters in multi-planet systems. Symbols are colored blue if the warm Jupiter orbits a metal-poor host ([Fe/H] $<$ 0) and orange if they orbit a metal-rich host ([Fe/H] $>$ 0). The grey shaded area shows tracks of constant angular momentum which planets follow during tidal circularization. Here we have adopted a mass of 1 $M_\mathrm{Jup}$ and a  radius of 1 $R_\mathrm{Jup}$ around a Sun-like star. The dashed black line displays where $f$-mode tidal dissipation could speed up tidal migration (\citealt{Wu2018}; \citealt{Dong2021}).} 
\label{fig:a_e}
\end{center}
\end{figure*}

The remainder of this paper is organized as follows. In Sections \ref{sec:Target_Sample} and \ref{sec:Host_Stars} we discuss our warm Jupiter target selection and their stellar hosts. We describe the HBM framework in Section \ref{sec:HBM}. In Section \ref{sec:Results} we contextualize our results and our interpretation in the framework of giant planet migration. Possible interpretations of the impact of stellar host mass and metallicity on the underlying eccentricity distributions of warm Jupiters is discussed in Section \ref{sec:Discussion}. Finally, an overview of our conclusions can be found in Section \ref{sec:Conclusion}.

\section{Target Sample and Orbit Fits}\label{sec:Target_Sample}

\subsection{ Overview of Sample and Orbit Fits}\label{sec:orbit_fits}
This homogeneous reanalysis of warm Jupiter eccentricities initiated with the selection of planets from the NASA Exoplanet Archive, as of May 2022, (\citealt{Akeson2013}) with orbital periods between 10-365 days ($\approx$0.1-1 AU), and a measured minimum mass of $m_p \sin i$ = 0.3--13 $M_\mathrm{Jup}$, or a radius $>$8 ${\rm R}_{\oplus}$. We exclude transiting planets that have not also been recovered with RVs as they cannot be used in our orbit refits. In addition, we exclude hot Jupiters with periods $<$10 days ($\approx$0.1 AU)  because their orbits have most likely been shaped by tidal circularization (\citealt{RasioTides1996}; \citealt{Lubow1997}).

We use the \texttt{RadVel} package (\citealt{Fulton2018}), which utilizes the Markov chain Monte Carlo (MCMC) sampling routine \texttt{emcee} (\citealt{Foreman-Mackey2013}), to model the radial velocity time series data and re-fit the Keplerian orbits of all warm Jupiters in our sample. In Paper I (Morgan et al. 2025, in press) we provide details of our sample selection, orbit refits, and the finalized sample of 200 warm Jupiters orbiting 194 host stars. 

\subsection{Warm Jupiter Host Stars}\label{sec:Host_Stars}
In Paper I, we presented a compilation of stellar masses, metallicities, effective temperatures, and distances for the 194 host stars in the sample together with their associated uncertainties (Morgan et al. 2025, in press). These were assembled from a multitude of sources including discovery papers, follow-up studies cataloged in VizieR (\citealt{Ochsenbein2000}), the NASA Exoplanet Archive (\citealt{Akeson2013}; \citealt{Christiansen2025}), and Gaia DR3 (\citealt{GaiaCollaboration2023}; \citealt{BailerJones2021}).

\begin{figure*}
  {\includegraphics[width=\linewidth, height=1.9\textheight, keepaspectratio]{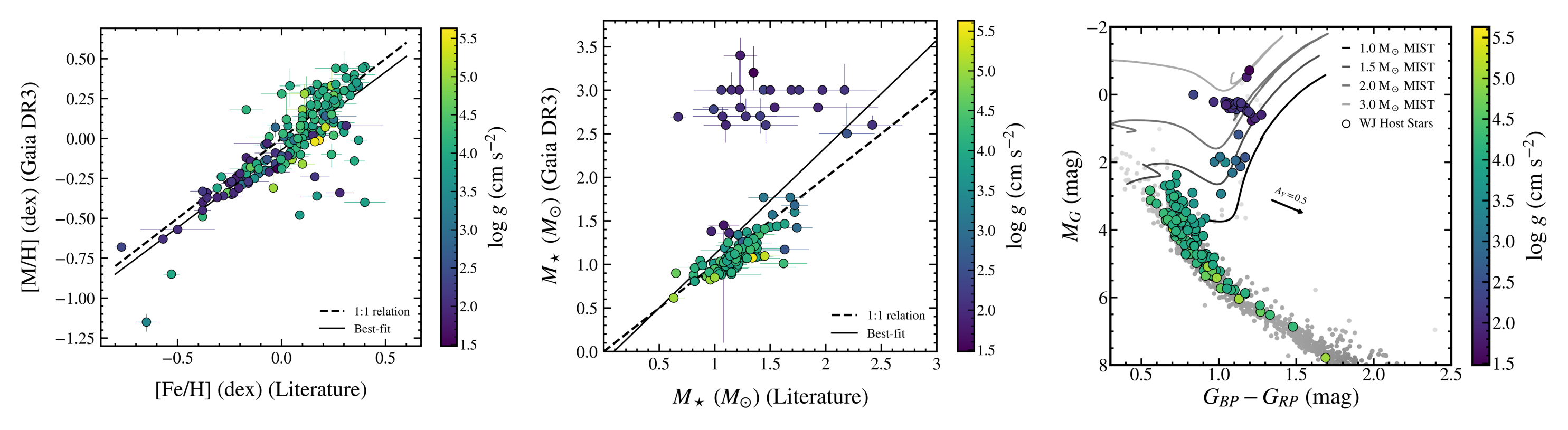}}
    \caption{Left: Stellar host metallicity measurements compiled from the literature and used in this analysis compared with metallicities derived from Gaia DR3 RVS spectra (\citealt{GaiaCollaboration2022}). Individual warm Jupiter host stars are colored by Gaia log $g$ to highlight main-sequence stars (log $g$ $>$ 4.0 dex) and post main-sequence stars (log $g$ $<$ 4.0 dex). The dashed line represents the 1:1 relation, while the solid line shows the best-fit linear relation. Middle: Stellar masses used in this analysis compared with uniformly inferred FLAME masses from Gaia. Systems that lie well above the 1:1 line have Gaia surface gravities suggesting they are massive post-main-sequence stars. Right: Host stars plotted on a Gaia $M_{G}$ vs $G_{BP}$ -- $G_{RP}$ color-magnitude diagram. Iso-mass tracks spanning 1.0--3.0 $M_\odot$ are shown with solar metallicities ([Fe/H] = 0 dex). The most evolved stars tend to have the highest stellar masses.}
    \label{fig:mass_met_comparisons}
\end{figure*}

Here, we provide a summary of host star characteristics to contextualize the comparative warm Jupiter eccentricity tests carried out later in this study. Most host stars in our survey are nearby (d $<$ 200 pc), bright ($V$ $<$ 10 mag) FGK stars with spectroscopically determined metallicities. The median host star metallicity is 0.09 dex with most metallicities (98$\%$) ranging from $-$0.50 to $+$0.50 dex. The median stellar metallicity uncertainty is 0.04 dex, with most under 0.1 dex.

The majority of the mass measurements and uncertainties in this analysis are derived from isochrone fitting and represent reported values from discovery papers or follow-up characterization studies. The median host star mass is 1.13 $M_{\odot}$ with most masses ranging from about 0.5--2 $M_{\odot}$. The median stellar mass uncertainty is 0.08 $M_{\odot}$ with most uncertainties under 0.3 $M_{\odot}$. This precision is typical of mass estimates for bright stars with Gaia DR3 parallaxes (\citealt{Serenelli2021}).

 The host star properties in this work are derived from a heterogeneous array of techniques and observations spanning several decades. The advantage of this approach  is that, in general, more care was taken to determine the mass of a particular star hosting a new planet. On the other hand, this compilation could suffer from systematic biases related to the variety of methodologies. We test the consistency of the mass and metallicity measurements used in this analysis by comparing them to the stellar masses and metallicities derived from Gaia's Radial Velocity Spectrometer (RVS) spectra (\emph{R}$\approx$11,500) which is calibrated and analyzed by the Gaia Generalized Stellar Parameterizer-Spectroscopy module of the Gaia Astrophysical Parameters Inference System pipeline (\citealt{GaiaCollaboration2022}). In particular, we compile the metallicity, [M/H], and Final Luminosity Age Mass Estimator (FLAME) masses; among 194 host stars, 180 have metallicities and 138 have masses (Figure \ref{fig:mass_met_comparisons} left and middle panels). 

We find that the  measurements are generally in good agreement with each other as seen in Figure \ref{fig:mass_met_comparisons}. A maximum likelihood estimation was performed using a simple linear model of the form $\mathrm{[M/H]}_\mathrm{DR3}$ = $a_1$ + $b_1$$\mathrm{[Fe/H]}_\mathrm{lit}$ for metallicity and $M_{\star,\mathrm{DR3}}$ = $a_2$ + $b_2$ $M_{\star,\mathrm{Lit}}$ for stellar mass, where  $M_{\star,\mathrm{DR3}}$ and $M_{\star,\mathrm{Lit}}$ are in units of solar masses. We find that for the metallicity relation, the best-fit parameters are $a_1 = -0.07 \pm 0.01$ and $b_1 = 0.96 \pm 0.15$. For the stellar mass relation, the best-fit parameters are $a_2 = -0.11 \pm 0.22$ and $b_2 = 1.23 \pm 0.59$.

The few systems with less precise metallicity measurements tend to be fainter, more distant stars. In the stellar mass comparison plot (middle panel in Figure \ref{fig:mass_met_comparisons}), there is a cluster of stars that Gaia implies are high mass (2.5--3.5 $M_{\odot}$)  but which have lower mass estimates (1.0--2.0 $M_{\odot}$) from our compilation of literature values. These stars have larger uncertainties in stellar mass and are generally associated with what the Gaia log $g$ values (1.5--3.0 cm s$^{-2}$) suggest are evolved stars.

To further analyze this discrepancy, we plot the stars with FLAME masses on a Gaia color-magnitude diagram (CMD) in the right panel of Figure \ref{fig:mass_met_comparisons} (\citealt{Riello2021}). Stars are de-reddened using extinction values reported in the General Stellar Parameterizer from Photometry (GSP-Phot) catalog (\citealt{GaiaCollaboration2022}; \citealt{Andrae2023}). MESA Isochrones $\&$ Stellar Tracks (MIST) models (\citealt{Paxton2011}; \citealt{Choi2016}; \citealt{Dotter2016}), computed with synthetic Gaia EDR3 photometry and assuming solar metallicity, are overplotted for stars spanning a range of masses. The evolved stars most discrepant in mass with our catalog seem to fall between the 2.0--3.0 $M_{\odot}$ iso-mass tracks, supporting the Gaia-determined mass. However, degeneracies in isochrone fitting, uncertainties in the metallicity determinations, and uncertainties in optical line lists and opacities at late types can impact age and mass estimates. This likely explains the discrepancy between values reported in discovery papers and those inferred from Gaia. A more rigorous stellar host characterization is likely needed to establish which parameters are more reliable. Nevertheless, the impact on our final results in Section \ref{sec:Host_Star_Mass} (below) are minimal as most of these evolved stars already reside in our high-mass subsample.

\subsection{Comparison to Solar Neighborhood and Galactic Trends}\label{sec:Solar_Neighborhood_Galactic_Comparison}

\begin{figure}
\begin{center}
{\includegraphics[width=\linewidth]{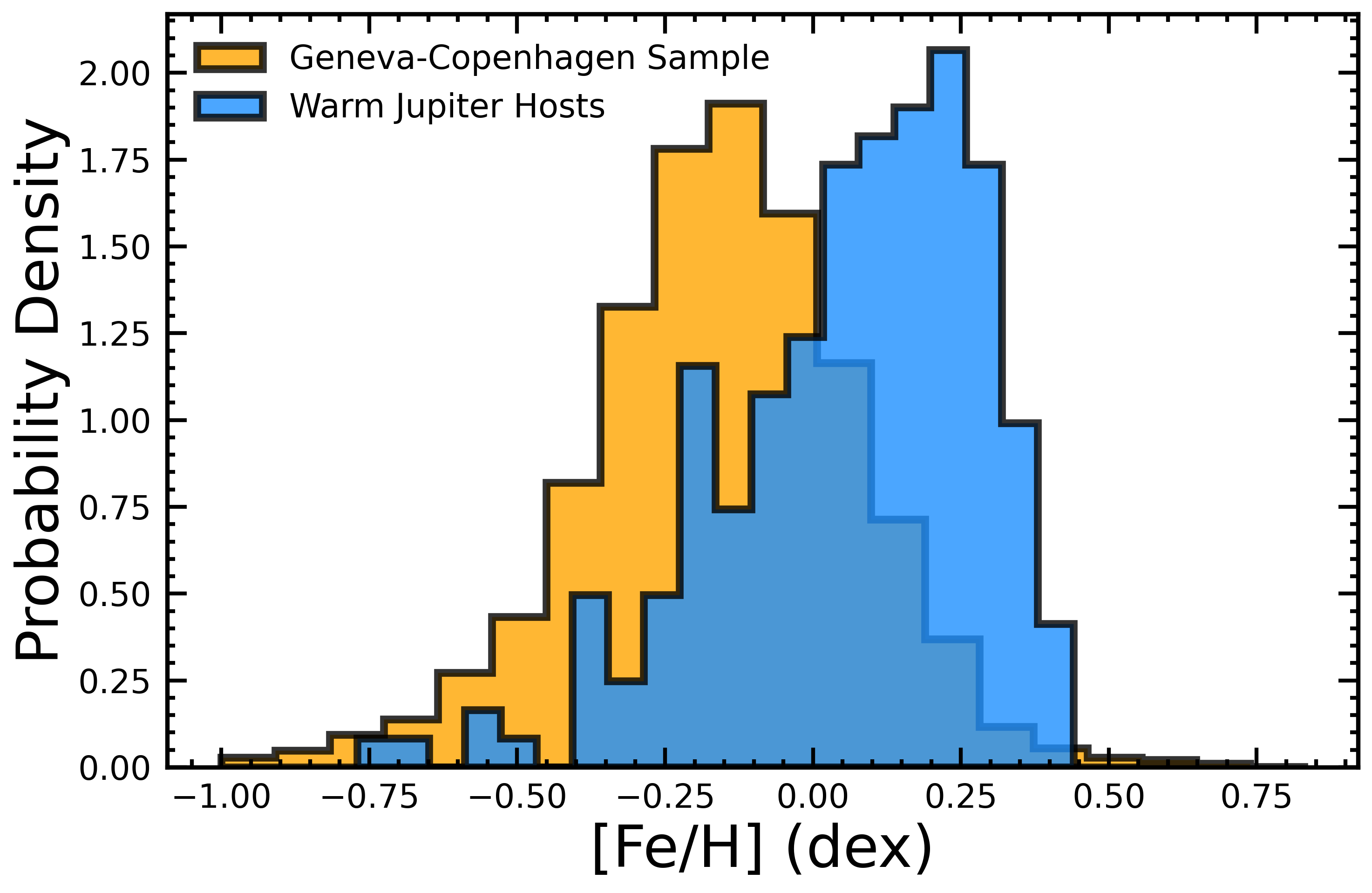}}
\caption{ Plotted in orange is the metallicity distribution of $\approx$20,000 field stars in the solar neighborhood within $\sim$40 pc (\citealt{Nordstrom2004}). The blue histogram shows the metallicity distribution of warm Jupiter host stars in this study. These hosts are, on average, more metal-rich than the field star population.}
\label{fig:metallicity_GC_comparison}
\end{center}
\end{figure}

How representative is the sample of warm Jupiter hosts analyzed in this study compared to nearby stars? There is a long history of volume-limited spectroscopic analyses of stars in the solar neighborhood to determine metallicities and kinematics. For instance, \citet{Nordstrom2004}, \citet{Luck2005}, \citet{Gray2006}, and \citet{Casagrande2011} examined tens of thousands of stars within 40 pc and found that a substantial fraction of nearby field stars have subsolar ($\langle \text{[Fe/H]} \rangle \approx -0.10$ dex) metallicities, with the distribution encompassing both metal-poor and metal-rich stars. Notably, there is a significant deficit of metal-poor Sun-like stars (\citealt{Jorgensen2000}), which is relevant since most stars in RV surveys are nearby FGK dwarfs. In our analysis, the average metallicity of host stars is $\langle \text{[Fe/H]} \rangle = 0.09$ dex, or about 0.2 dex higher than that for a similar nearby star that does not have any prior information about its planetary system. The distribution of metallicities in this subsample for stars hosting warm Jupiters is therefore not representative of the typical values in the solar neighborhood as there is a selection bias favoring metal-rich systems (see Figure \ref{fig:metallicity_GC_comparison}). This is in line with the known giant planet-metallicity relation (\citealt{Santos2004}; \citealt{FischerValenti2005}; \citealt{Johnson2010}).

\begin{figure*}
  \centering
    \begin{minipage}[b]{0.48\textwidth}
        \centering
        \includegraphics[width=\linewidth]{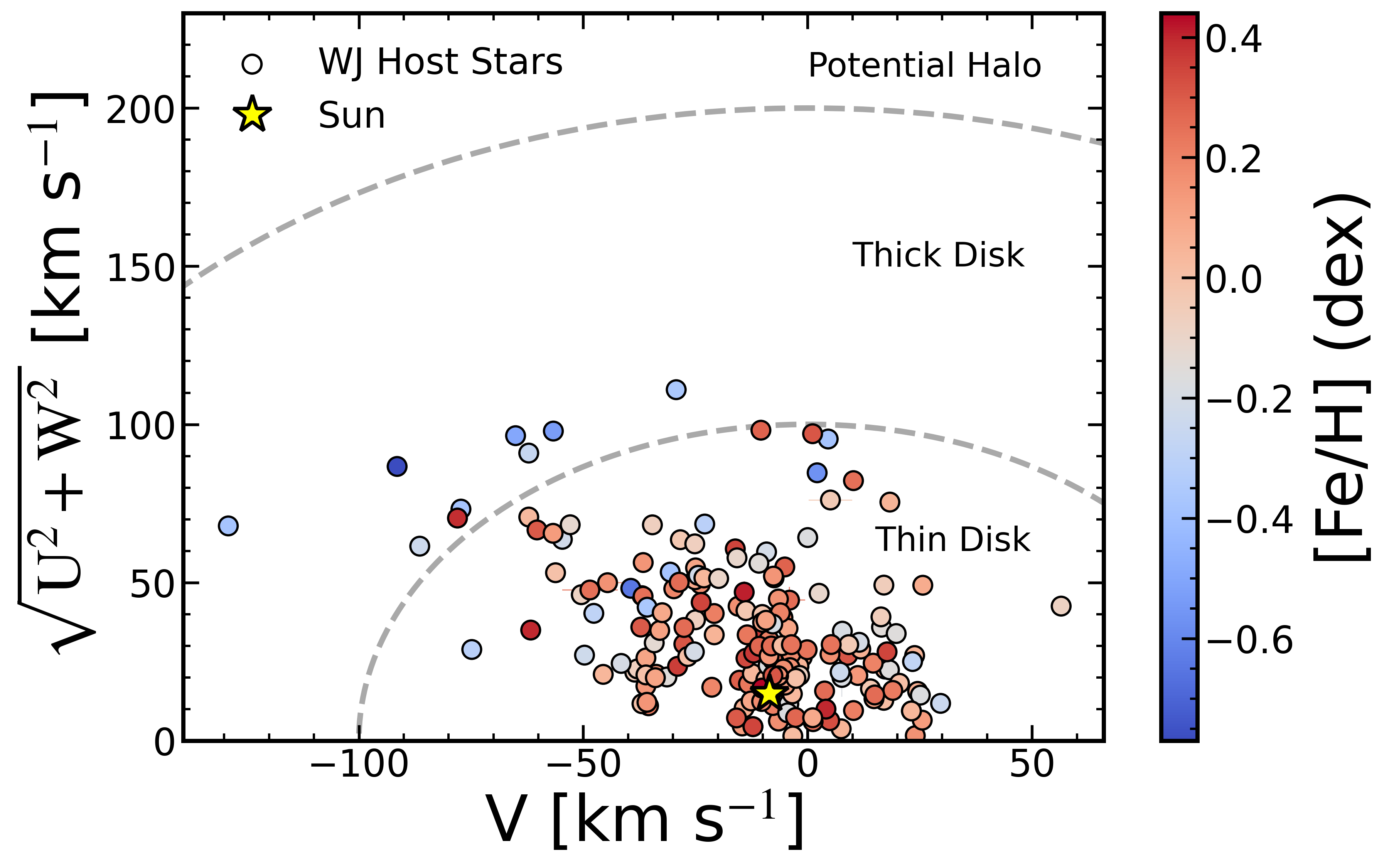}
    \end{minipage}
    \hfill
    \begin{minipage}[b]{0.48\textwidth}
        \centering
        \includegraphics[width=\linewidth]{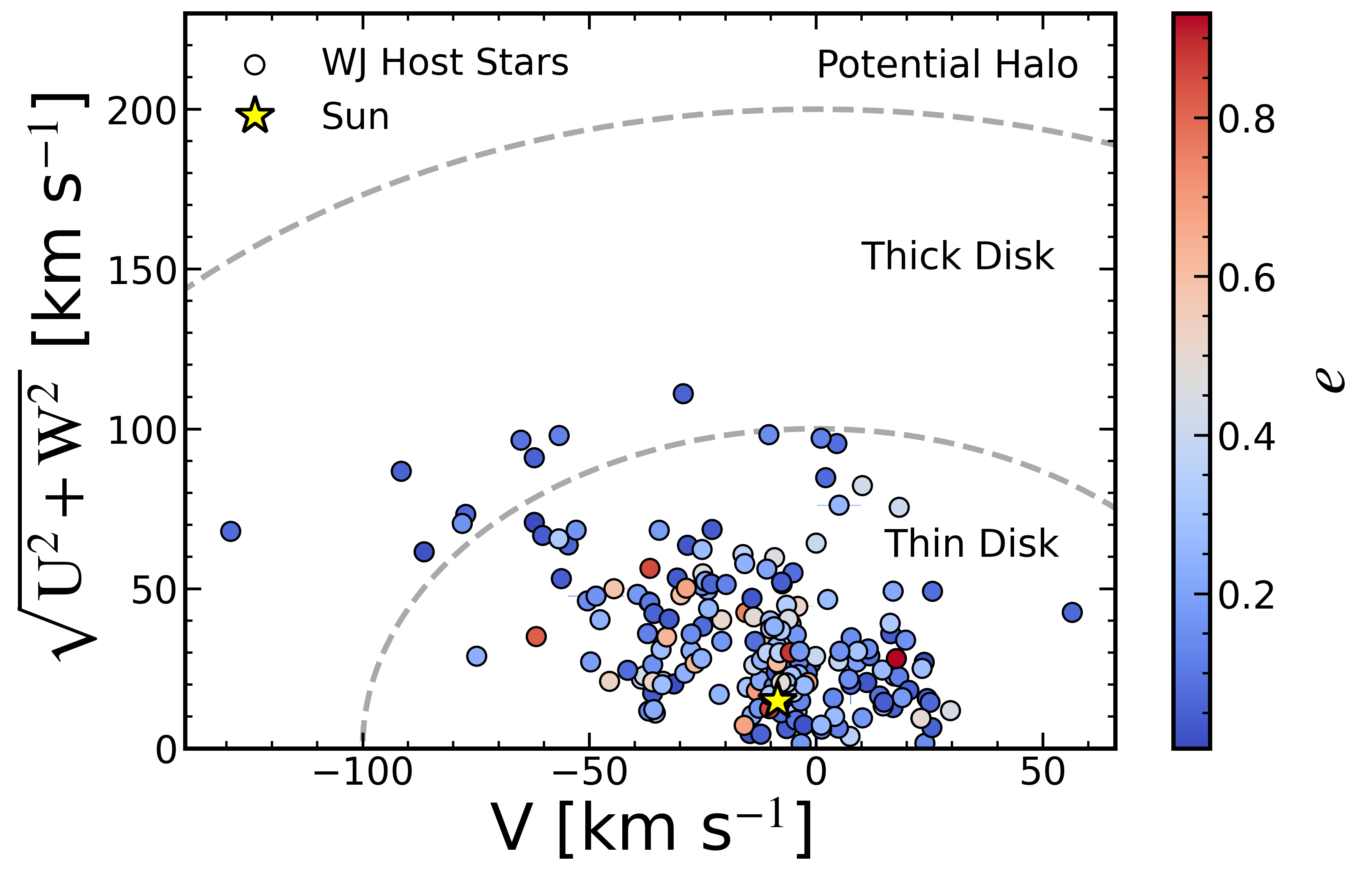}
    \end{minipage}
    \caption{Toomre diagrams illustrating $\mathit{U},\mathit{V},\mathit{W}$ Galactic kinematic space velocities of the warm Jupiter host stars in this analysis using their Gaia DR3 measurements. Contours separate stars that are kinematically most similar to the thin disk, thick disk, and halo (see \citealt{Bensby2003}). We only plot systems with kinematic uncertainties less than 20$\%$. Left: Here, symbol colors relate to stellar metallicity, showing a trend in which the most metal-poor host stars in our sample belong preferentially to the galactic thick disk. Right: Here, symbol colors scale with the eccentricity of the warm Jupiter. All warm Jupiters orbiting thick disk stars have near-circular orbits.}
    \label{fig:toomre}
\end{figure*}

Stellar properties such as distance, kinematics, and age can correlate with a star’s metallicity, potentially impacting the formation and evolutionary pathways of the giant planets they may harbor (e.g. \citealt{Maldonado2018}; \citealt{Banerjee2024}). For instance, \citet{HamerSchlaufman2019} found that hot Jupiter host stars are, on average, younger than the field star population, using Galactic velocity dispersions which correlate with stellar age. Here we assess whether warm Jupiter hosts belong to a mixture of galactic populations or are predominantly members of the thin disk, as might be expected given that most systems in our sample (80$\%$) are located within 200 parsecs. Using the Python package \texttt{PyAstronomy} (\citealt{Czesla2019}) we calculate the Galactic positions and space velocities using Gaia DR3 proper motions, parallaxes, and radial velocities. $\mathit{U}\mathit{V}\mathit{W}$ velocities have been corrected for the Solar motion relative to the local standard of rest by adopting values of ($\mathit{U}_\odot, \mathit{V}_\odot, \mathit{W}_\odot$) = (8.5, 13.38, 6.49) km s$^{-1}$ from \citet{Coșkunoglu2011}.\footnote{$\mathit{U}$ points radially inward toward the Galactic center, $\mathit{V}$ is directed along the direction of Galactic rotation, and $\mathit{W}$ is directed upward towards the Galactic North pole.} To account for velocity uncertainties, we perform Monte Carlo realizations of $\mathit{U},\mathit{V},\mathit{W}$ velocities assuming normal distributions for the input parameters. The results are shown in the Toomre diagram in Figure \ref{fig:toomre}, color-coded by host metallicity. 

The majority of stars in this sample (95$\%$) are consistent with being members of the Milky Way thin disk while the remaining 5$\%$ are kinematically similar to thick disk stars. Our sample is therefore not representative of the typical kinematics of field stars. \citet{Bensby2014} analyze the kinematics of 714 Sun-like stars in the solar neighborhood (within 100 pc) and found that $\sim$$54\%$ are most likely thin disk stars, $\sim$$33\%$ exhibit kinematics consistent with the thick disk, and $\sim$$13\%$ have intermediate kinematics that are indistinguishable between the two populations. 

While we lack complete information about the parent samples from the surveys that were conducted to find giant planets, the warm Jupiters that were discovered have been found orbiting stars that are more metal-rich and kinematically consistent with the thin disk population. 

\section{Hierarchical Bayesian Modeling}\label{sec:HBM}

 We use HBM to fit the underlying behavior of the warm Jupiter population. HBM is a statistical framework that simultaneously infers parameters of individual systems and hyperparameters governing a parent population. Here, we follow the statistical framework outlined in \citet{Hogg2010}, which has also been used to study the eccentricity distributions of small transiting planets (\citealt{VanEylen2019}) and directly imaged giant planets (\citealt{Bowler2020}). This sampling framework was incorporated into \texttt{ePop!} (\citealt{Nagpal2023}), an open source Python package for fitting population-level distributions to sets of individual system distributions, assuming a parametric model for the underlying eccentricity distribution. 

Our general strategy is to subdivide targets into two subsamples and then compare the inferred underlying distribution functions.  One of these models we test in this work is the Beta distribution,
\begin{equation}
    \label{beta_equation}
      B(\epsilon)=\frac{\Gamma(\alpha+\beta)}{\Gamma(\alpha) \Gamma(\beta)}\epsilon^{\alpha-1}(1-\epsilon)^{\beta-1},
\end{equation}
a flexible model with two shape parameters $\alpha$ and $\beta$.\footnote{Note that we use $\epsilon$ in Equation 1 for eccentricity to avoid ambiguity with the Euler's number.}  Hyperpriors must be chosen for $\alpha$ and $\beta$; we test the impact of two hyperpriors (assumed to be equal for $\alpha$ and $\beta$): a truncated Gaussian, and a log-Uniform with $\mu$ = 0.69 and $\sigma$ = 1.0 (see Table 1 in \citealt{Nagpal2023}). The Beta distribution has been found to to be a reasonable representation for the underlying eccentricity distribution of all exoplanets detected through radial velocities (\citealt{Kipping2013}).

We randomly draw 10$^{3}$ samples for each planet eccentricity distribution to ensure computational Tractibility for the HBM analysis. We then ran 40 walkers for 4 $\times$ $10^{3}$ steps and include a burn in fraction of 50$\%$. Markov chain Monte Carlo (MCMC) convergence using \texttt{emcee} (\citealt{Foreman-Mackey2013}) is assessed from the Gelman-Rubin statistic, trace plots, and corner plots (\citealt{Foreman-Mackey2016}). The joint posterior distributions between the hyperparameters can be found in Appendix \ref{sec:joint_posterior_distributions} (e.g. Figure \ref{fig:corner_plots}).

\begin{figure}
  \centering
  
  \begin{minipage}[b]{0.45\textwidth}
    \centering
    \includegraphics[width=\linewidth]{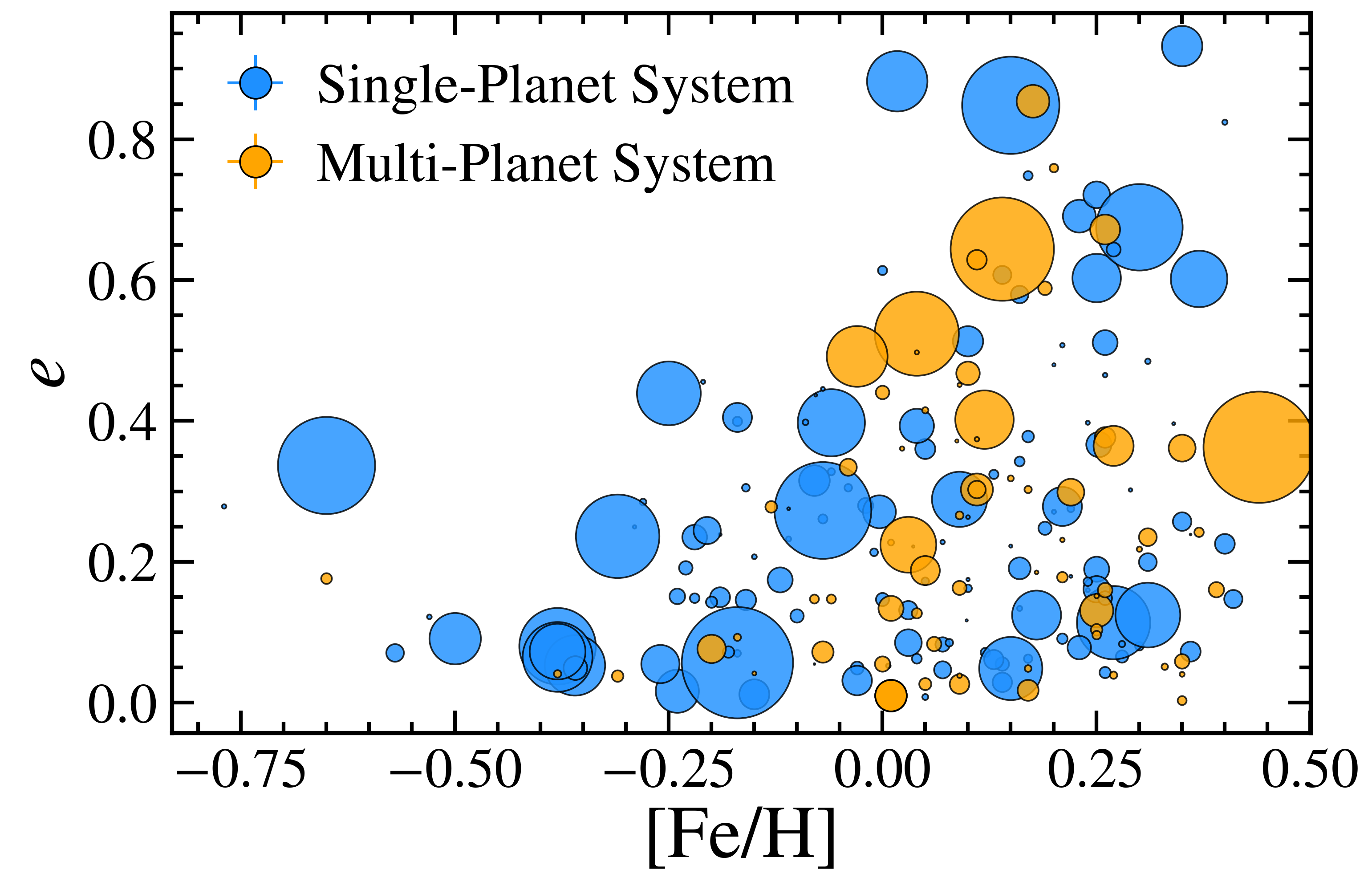}
  \hfill
    \centering
    \includegraphics[width=\linewidth]{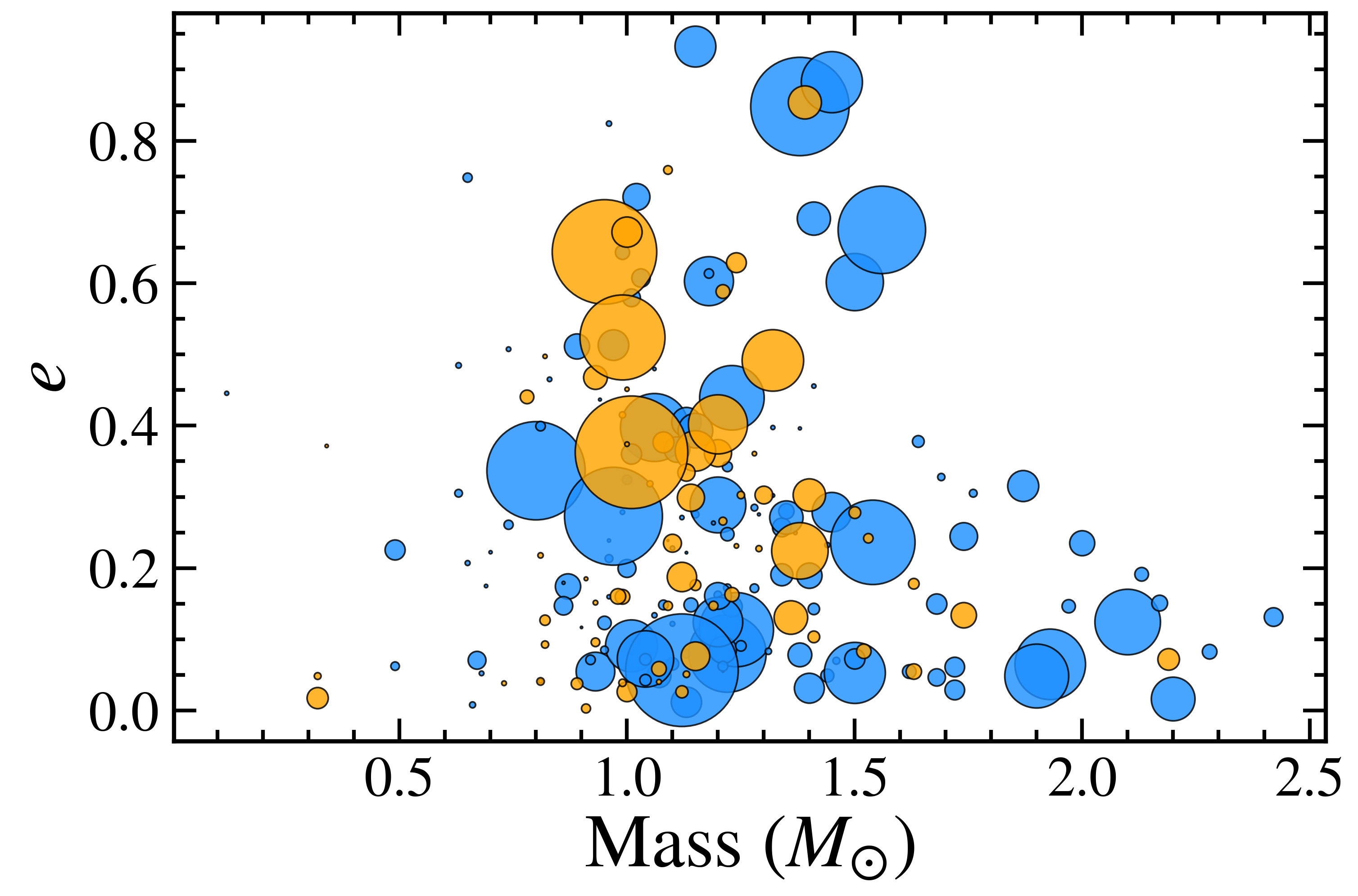}
    \hfill
    \centering
    \includegraphics[width=\linewidth]{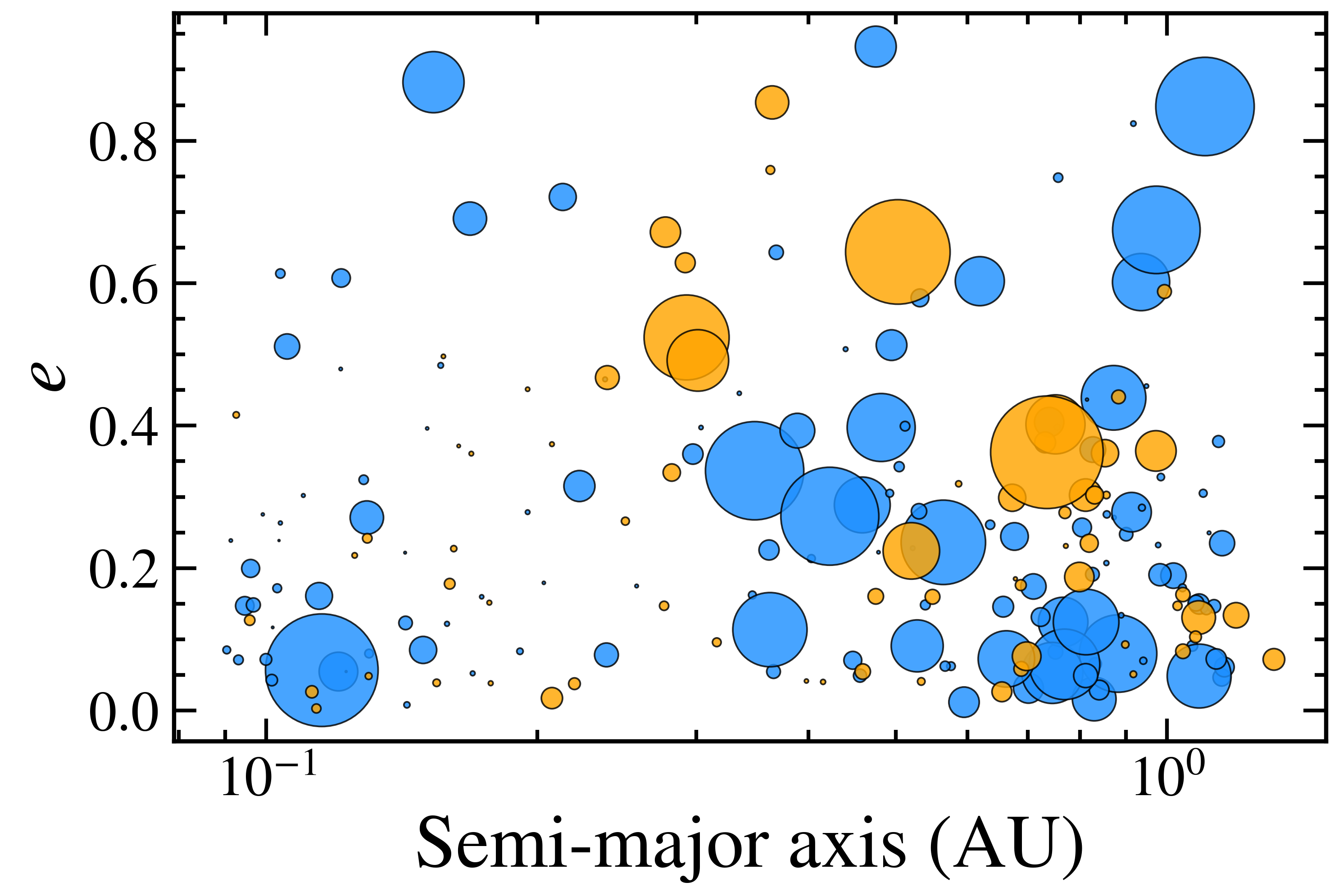}
    \caption{Top: Host star metallicity as a function of planet eccentricity for all warm Jupiters in single planet systems (blue) and multi-planet systems (orange). Middle: Stellar host mass as a function of eccentricity for all warm Jupiters. Bottom: Semi-major axis as a function of eccentricity for all warm Jupiters. For all panels, the size of the circles scale with minimum planet mass, $m_p \sin i$.}
  \label{fig:planet_properties}
  \end{minipage}
\end{figure}

The Rayleigh distribution has been found to be a good fit to various orbital elements for a variety of ensembles of objects impacted by long-term dynamical evolution including the eccentricities of small planets in Kepler multi-planet systems (\citealt{VanEylenAlbrecht2015}) and the orbital inclinations of Main-belt Asteroids (\citealt{LiuWu2023}). In addition to the Beta distribution we implement a Rayleigh distribution,
 \begin{equation}
    R(\epsilon) = \frac{\epsilon}{\sigma^2} e^{-\frac{\epsilon^2}{2\sigma^2}},
\end{equation}
where $\sigma$ is bounded between 0--1.
A Gaussian distribution with both $\sigma$ and $\mu$ bounded between 0--1 is also tested:
 \begin{equation}
    G(\epsilon) = \frac{1}{\sigma \sqrt{2\pi}} e^{-\frac{(\epsilon - \mu)^2}{2\sigma^2}} .
\end{equation}
 
 Both distributions are modeled using a truncated Gaussian and a log-Uniform hyperprior. The Rayleigh and Gaussian distributions are tested to compare outcomes with the more flexible Beta distribution model.  Results are discussed in more detail in Appendix \ref{sec:Additional_Statistical_Tests}. 

\subsection{Accounting for Eccentricity Biases}\label{sec:Eccentricity_Biases}

The likelihood model used in the HBM framework is given by,
\begin{equation}
\mathcal{L_{\mathit{v}}} = \prod_{i=1}^{N} \frac{1}{K_i} \sum_{j=1}^{K_i} B_{\mathit{v}}(e_{ij}),
\end{equation}
where $N$ is the number of systems, ${K_i}$ is the number of samples contained within the $i$th individual
eccentricity posterior, $e_{ij}$ is the $j$th eccentricity sample for the $i$th system, and \textit{v} is the vector of parameters in the population-level model (\citealt{Nagpal2023}). However, this model does not account for potential biases associated with planets discovered with the RV and transit methods. These biases can potentially affect the shape of the underlying warm Jupiter eccentricity distribution. This analysis includes 155 planets (78$\%$) discovered through RVs and 45 planets (22$\%$) discovered via the transit method.
  
\citet{ShenTurner2008} found that RV-detected planet eccentricities can be overestimated in the regime where velocity semi-amplitudes are low compared to RV uncertainties. This Keplerian fitting bias tends to inflate the inferred eccentricities of intrinsically low-eccentricity orbits, thereby altering the observed eccentricity distribution for a sample of planets. They demonstrate that this effect becomes most relevant for RV-detected systems when 
\begin{equation}
\left( \frac{K_{\mathrm{fit}}}{\sigma} \right) \sqrt{N} < 17,
\label{eq:kfit_threshold}
\end{equation}
where $K_{\mathrm{fit}}$ is the fitted semi-amplitude, $\sigma$ is the radial velocity uncertainty, and $N$ is the number of observations. This bias affected only 10$\%$ of the sample in their analysis, and they conclude that it would become significant if more than 10$\%$ of a sample of RV planets were impacted. In our analysis only $\sim$7$\%$ (11/155) of the RV-discovered planets satisfy Equation \ref{eq:kfit_threshold}. Therefore, the orbit-fitting bias is not likely to be significantly impacting the underlying eccentricity distribution and we therefore do not make corrections for the subsample of RV-discovered planets.

For transiting planets, there is a geometric bias that favors more eccentric planet discoveries due to an increased transit probability (\citealt{Barnes2007}; \citealt{Burke2008}; \citealt{Winn2010W}; \citealt{Kipping2014}). To account for this, a $(1 - e^2) / (1 + e \sin\omega)$ weighting factor can be applied to the likelihood which adjusts for the increased probability of observing more eccentric orbits. Using Equation 5 from \citet{Gilbert2025}, 

\begin{equation}
\mathcal{L_{\mathit{v}}} = \prod_{i=1}^{N} \frac{1}{K_i} \sum_{j=1}^{K_i} B_{\mathit{v}}(e_{ij}) \left( \frac{1 - e_{ij}^2}{1 + e_{ij} \sin \omega_{ij}} \right),
\end{equation}

\noindent we update the likelihood function to account for the over-representation of eccentric planets in transit surveys, down-weighting this observational bias in the posterior samples of transiting planets (also see \citealt{DongHuangTESS2021}; \citealt{Fairnington2025}). After incorporating this correction factor into our likelihood, we found that that it only subtly impacts the inferred eccentricity distribution, but we adopt this transit bias-corrected version for all of the tests in this study.

\section{Results}\label{sec:Results}

\subsection{The Warm Jupiter Eccentricity Distribution}\label{sec:Previous Results}

\begin{figure}
\begin{center}
{\includegraphics[width=\linewidth]{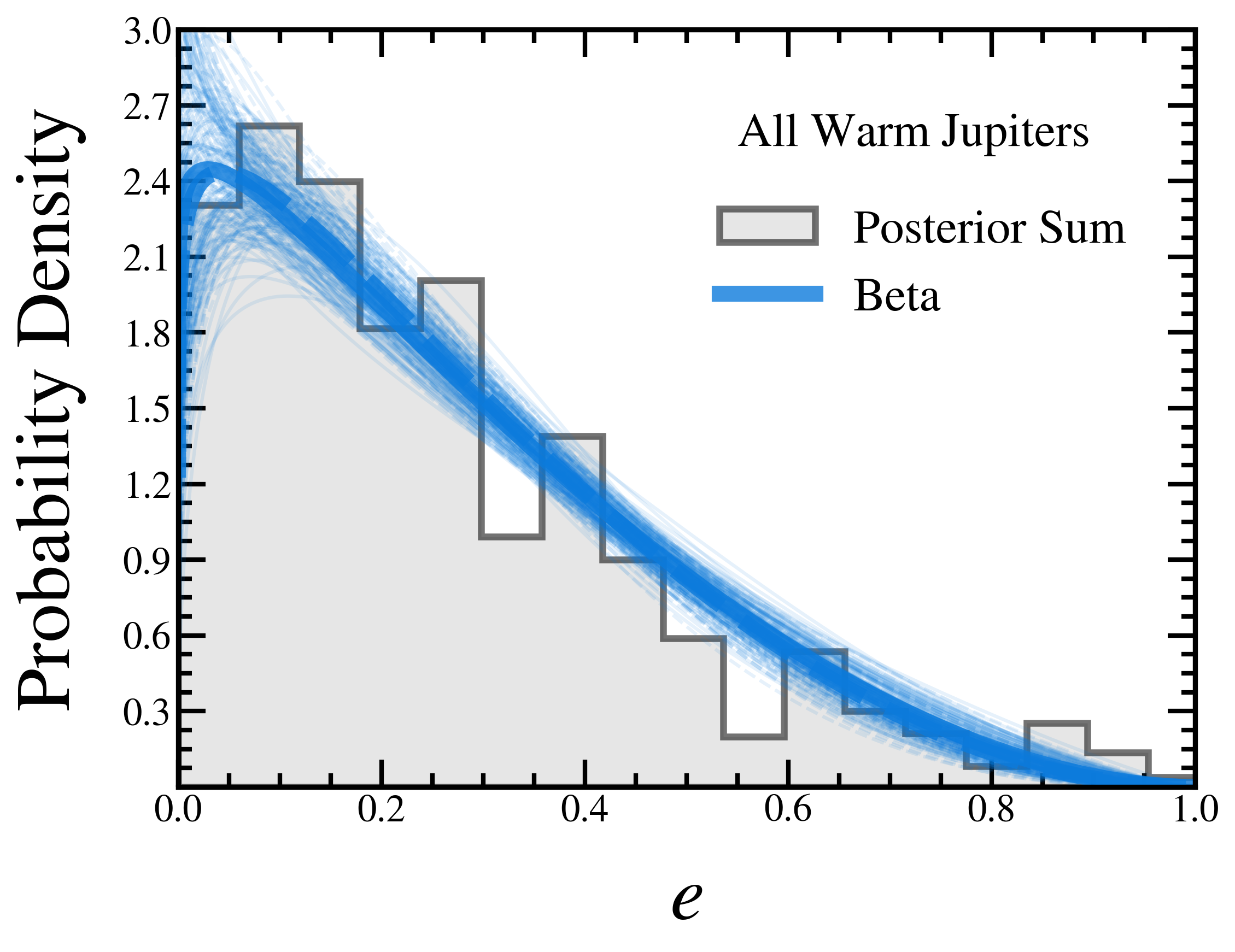}}
\caption{Underlying eccentricity distribution results from our hierarchical Bayesian modeling for all warm Jupiters.  The blue curve displays the median inferred underlying $\mathit{e}$ distributions modeled by the Beta distribution. Solid lines represent a Truncated Gaussian hyperprior while the dashed lines represent a log-Uniform hyperprior. Thin lines in the background show randomly sampled distributions from the  $\alpha$ and $\beta$ hyperparameter posteriors. The summation of eccentricity posteriors are plotted as a histogram in grey.} 
\label{fig:True_WJ_Distribution}
\end{center}
\end{figure}

We begin by modeling the inferred underlying eccentricity distribution for all warm Jupiters in the sample.  Results for the Beta distribution model are displayed in Figure \ref{fig:True_WJ_Distribution}.  The full warm Jupiter eccentricity distribution yields $\alpha$ and $\beta$ values of 1.00$^{+0.09}_{-0.08}$ and 2.79$^{+0.28}_{-0.26}$, respectively for the Beta distribution. This can serve as a reasonable eccentricity prior for all future warm Jupiter orbital fits. For the Rayleigh distribution we find a spread of $\sigma_{R}$ = 0.23$^{+0.01}_{-0.01}$, and the Gaussian distribution yields $\mu_{G}$ = 0.26$^{+0.01}_{-0.01}$ and $\sigma_{G}$ = 0.20$^{+0.01}_{-0.01}$ (see Appendix \ref{sec:Additional_Statistical_Tests}). The general shape of the inferred distributions is similar: Warm Jupiters span a wide range of eccentricities, with a preference for low values but substantial power out to high eccentricities of \emph{e} $\sim$ 0.8. 27$^{+3}_{-4}\%$ of warm Jupiters recovered from RVs have eccentricities consistent with near-circular orbits ($e$ $<$ 0.1), suggesting that most warm Jupiters (73$^{+3}_{-3}\%$) detected are dynamically hot.

Our results are in good agreement with previous attempts to infer the eccentricity distribution of warm Jupiters. \citet{DongHuangTESS2021} fit the eccentricity distribution of 55 warm Jupiters discovered in the first year of the TESS Full-frame Images using various parametric models. They found $\alpha$ = 1.78$^{+1.39}_{-0.77}$ and $\beta$ = 4.08$^{+2.65}_{-1.63}$ for the Beta distribution model and a $\sigma_{R}$ = 0.259$^{+0.032}_{-0.029}$ from the Rayleigh distribution. Our results are also consistent with \citet{Rosenthal2023}, for which they report $\alpha$ = 0.85$^{+0.13}_{-0.11}$ and $\beta$ = 2.27$^{+0.36}_{-0.32}$ for 80 giant planets beyond 0.3 AU in The California Legacy Survey. Similarly, \citet{KaneWittenmyer2024} find a median eccentricity of 0.24 for 651 giant planets oribiting interior to the snow line (excluding hot Jupiters). 

\subsection{Metal Rich Host Stars Harbor More Eccentric Warm Giants}\label{sec:Host_Star_Metallicity}
Correlations between host star metallicity and planet properties provide insights into planet formation and migration. The occurrence of giant planets  has been found to positively correlate with host star metallicity (\citealt{Gonzalez1997}; \citealt{Santos2001}; \citealt{FischerValenti2005}). This is typically interpreted within the context of planet formation efficiency. Metal-rich protoplanetary disks are expected to enhance the formation of giant planet cores because a higher prevalence of solids in the disk will accelerate the timescale of core growth (see \citealt{Petigura2018}). More giant planets would lead to increased planet-planet gravitational interactions and excite eccentricities. \citet{Dawson2013ApJ} found that the most eccentric warm Jupiters orbit metal-rich stars, which seems to support this general picture. 

\begin{figure*}
\begin{center}
{\includegraphics[width=\linewidth, height=1.9\textheight, keepaspectratio]{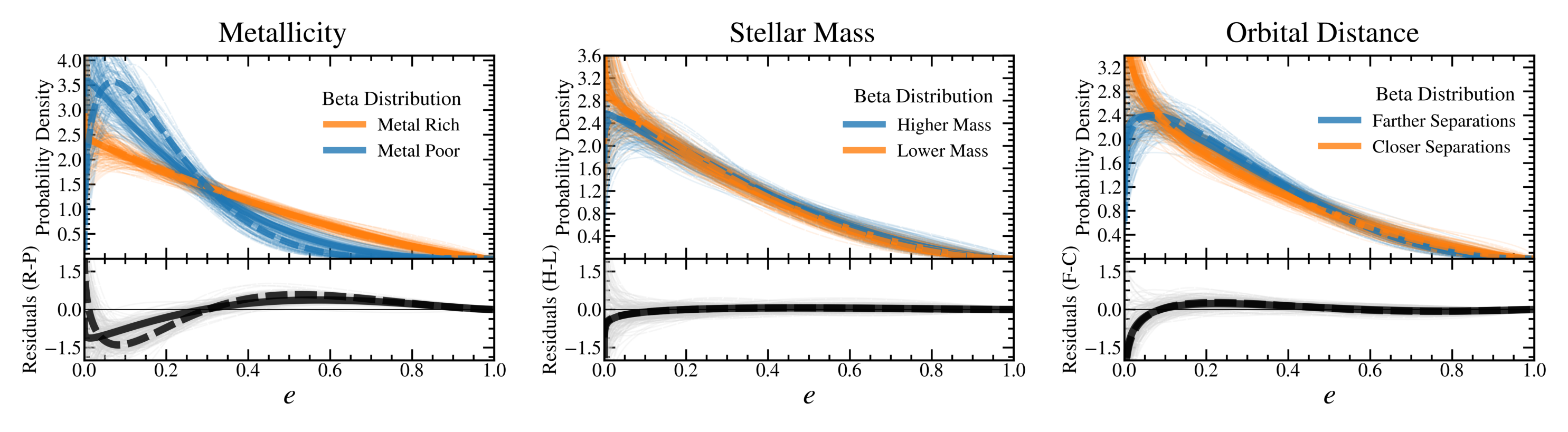}}
\caption{Results from our HBM test assuming a Beta distribution for the underlying eccentricity distribution with an emphasis on host star metallicity, mass, and orbital separation. Left: The orange and blue lines display the median inferred underlying eccentricity distribution of warm Jupiters when considering only the impact of stellar metallicity. Solid lines represent a truncated Gaussian hyperprior while the dashed lines represents a log-Uniform hyperprior. The lower panel displays the residuals of planets orbiting metal-rich host stars ([Fe/H] $\geq$ 0) subtracted from planets residing in metal-poor systems ([Fe/H] $<$ 0). Thin gray lines in the background show randomly sampled distributions from the $\alpha$ and $\beta$ hyperparameter posteriors. Middle: HBM test analyzing host star mass. The lower panel shows the residuals of the distribution of planets around higher-mass host stars with $M_{\star}$ $\geq$ 1.15 $M_{\odot}$ subtracted from lower-mass host stars with $M_{\star}$ $<$ 1.15 $M_{\odot}$. Right: HBM test with a focus on orbital separation. The lower panel shows the residuals of the distribution of planets at farther separations (100 d $\le$ $\mathit{P}$ $<$ 365 d) subtracted from those at close separations from their host stars (10 d $<$ $\mathit{P}$ $<$ 100 d).} 
\label{fig:beta_distributions}
\end{center}
\end{figure*}

Among the three parameters we analyzed---host star metallicity, stellar host mass, and orbital separation---the most significant difference was found for host star metallciity. In our analysis, the inferred eccentricity distributions of warm Jupiters around metal-rich and metal-poor stars differ from each other most notably at high eccentricities between 0.6 $<$ \emph{e} $<$ 0.93. Using solar metallicity as the cutoff, we compared 136 planets orbiting metal-rich host stars ([Fe/H] $\geq$ 0) against 64 planets residing in metal-poor systems ([Fe/H] $<$ 0). As seen in Figure \ref{fig:beta_distributions}, the underlying eccentricity distributions of warm Jupiters as a function of host star metallicity show that the most eccentric planets orbit metal-rich stars. Irrespective of underlying model and hyperprior (see Figure \ref{fig:additional_distributions1}), our tests reveal that eccentricities around metal-poor stars fall off around \emph{e} $\approx$ 0.60 while they span the full eccentricity range around metal rich stars. Our tests confirm the results found by \citet{Dawson2013ApJ}.

\subsection{Stellar Host Mass Has Minimal Effect on Warm Jupiter Migration}\label{sec:Host_Star_Mass}
RV, transit, and direct imaging surveys of low-mass stars have shown that they lack Jupiter analogs in comparison to their higher-mass counterparts (\citealt{Endl2006}; \citealt{Bowler2015}; \citealt{Bryant2023}; \citealt{Gan2023}; \citealt{Pass2023}). Indeed, the giant planet occurrence rate has been shown to positively correlate with stellar host mass (\citealt{Johnson2007}; \citealt{Johnson2010}). A large protoplanetary disk mass will increase the abundance of solids readily available for core accretion, which can explain the higher prevalence of giant planets around intermediate- and high-mass stars. Moreover, $P$ $\propto$ $M_*^{-1/2}$, so for a given stellar mass and at a given separation, the orbital timescale increases which should increase the rate of the collisional growth of solids. Both \citet{Andrews2013} and \citet{Pascucci2016} found that the relationship between the mass of a protoplanetary disk scales steeply with the mass of the host star. As expected with metal-rich disks, massive protoplanetary disks should amplify the formation of a greater number of giant planets (\citealt{Laughlin2004}), thereby leading to an increased occurrence rate of planet-planet scattering events and the emergence of a dynamically hot population of planets, including warm Jupiters.
 
When focusing only on stellar mass and marginalizing over other parameters that could impact planet eccentricity, we find that the eccentricities of warm Jupiters around lower-mass ($M_{\star}$ $<$ 1.15 $M_{\odot}$) and higher-mass ($M_{\star}$ $\geq$ 1.15 $M_{\odot}$) stars are mutually consistent. A cutoff of 1.15 $M_{\star}$ was chosen to create comparable bin sizes, with 102 planets orbiting lower-mass stars and 98 planets residing in higher-mass star systems. Figure \ref{fig:beta_distributions} shows that the distributions are similar between both samples of host stars, to within our ability to constrain them. This test is particularly interesting because our results contradict what is expected from correlations between host star mass, protoplanetary disk mass, and frequency of giant planets. Instead, we find that low and high-mass stars harbor warm Jupiters with similar underlying eccentricity distributions, suggesting that host star mass does not strongly influence giant planet dynamics, at least within $\sim$ 1 AU. 

\subsection{No Distinction with Orbital Separation}\label{sec:Orbital_Separation}
We tested several orbital period bins to assess whether there exists an overabundance of circular or highly eccentric orbits at specific orbital distances which could help disentangle clues about planet migration. A subset of the warm Jupiter population lies in the ``period valley" ($\sim$10--100 days) where the occurrence rate per log interval of giant planets dips (\citealt{Jones2003}; \citealt{Udry2003}; \citealt{Wittenmyer2010}; \citealt{Santerne2016}). Comparing 85 warm Jupiters orbiting in the period valley against 115 warm Jupiters exterior to this region may provide clues about giant planet eccentricities being damped or excited at favorable regions between 0.1--1 AU (\citealt{Murray1998}; \citealt{Petrovich2014}; \citealt{Anderson2020}), and  in particular whether disk migration or high eccentricity migration transports giant planets to distinct orbital distances.

We find that orbital separation has little effect on the underlying eccentricity distribution of warm Jupiters as shown in Figure \ref{fig:beta_distributions}. There is no significant difference when probing warm Jupiters with orbital periods 10 d $<$ $\mathit{P}$ $<$ 100 d and 100 d $\le$ $\mathit{P}$ $<$ 365 days. This indicates that warm Jupiters interior to 1 AU and exterior to the zone of tidal influence may be transported to their current locations via similar processes. 

\subsection{Stellar Host Membership}\label{sec:thick_thin_disk}

Stellar kinematics and birth location can correlate with a star’s metallicity (see Section \ref{sec:Solar_Neighborhood_Galactic_Comparison}), which in turn may impact the properties of its protoplanetary disk and any resulting giant planets. Studying correlations between warm Jupiter orbital properties and the galactic environments in which their host stars formed may offer clues about giant planet formation and migration (\citealt{Bashi2022}; \citealt{Chen2021}). Within our sample, 9 stars exhibit kinematics consistent with the thick disk, while 185 align more closely with the thin disk population of the Milky Way.

As hinted in Figure \ref{fig:toomre} (right panel), the eccentricities of warm Jupiters in the thick disk population appear to be lower than their counterparts in the thin disk: all have eccentricities under 0.17, and the average thick disk eccentricity ($\langle e_{\mathrm{thick}} \rangle$ = 0.07) is 68$\%$ lower than the average thin disk eccentricity ($\langle e_{\mathrm{thin}} \rangle$ = 0.22). Here, we examine whether this represents a statistically significant difference or may be a result of the modest sample of thick disk hosts.

We want to find the probability that all 9 systems consistent with the thick disk have eccentricities $e$ $<$ 0.2. We first calculate the probability of drawing an eccentricity in the range $0 \le e < 0.2$ from the full inferred warm Jupiter eccentricity distribution with $\alpha = 1.00$ and $\beta = 2.79$. By integrating the probability density function over this range, we obtain a value of 0.463. The probability that 9 planets randomly drawn from the sample of all warm Jupiters would have $e$ $<$ 0.2 is $0.463^{9}$, or 0.0010. This corresponds to a significance of $\sim$3.3$\sigma$ indicating that the probability of all warm Jupiters orbiting thick disk stars in this analysis having near-circular eccentricities purely by chance is small.

In addition, we compared the thick disk stars to a subset of thin disk stars using the HBM framework in the same way as was carried out in Section \ref{sec:HBM}. The bulk comparison of all thick versus all thin disk stars may be biased because of the preference for thick disk stars to be more metal-poor, and therefore have lower eccentricity WJs. To compare against stars in the thin disk, we randomly selected 36 stars matched to the 9 thick disk stars in stellar metallicity and mass—four times as many as in the thick disk sample. Of the nine thick disk stars, eight are metal-poor and one is metal-rich. As a result, each thin disk subset consisted of 32 stars with subsolar metallicities ([Fe/H] $<$ 0) and 4 stars with supersolar metallicities ([Fe/H] $\geq$ 0).

When comparing the nine thick disk stars to the randomly selected sample of 36 thin disk stars, we find results consistent with the previous test. The underlying eccentricity distribution of the planets orbiting thick disk stars does not extend beyond $e$ $\sim$ 0.2 while the eccentricity distribution around thin disk stars with similar stellar properties extends to higher eccentricities ($e$ $\sim$ 0.6). To test the robustness of this result, we randomly drew 3 new sets of thin disk stars and repeated the HBM analysis, again finding similar outcomes.

This is the first hint that warm Jupiters orbiting thick disk stars may be less dynamically excited than warm Jupiters orbiting stars in the thin disk. The smaller eccentricities observed for thick disk planets may reflect fewer dynamical interactions, possibly due to a lower occurrence rate of multiple giant planets or a reduced frequency of close-in stellar binaries. \citet{HallattLee2025} predict that there should be a deficit in giant planet occurrence around thick-disk stars due to the time and environments in which the stars formed.

\begin{deluxetable*}{lcccccc}
\renewcommand\arraystretch{0.7}
\tabletypesize{\scriptsize}
\setlength{ \tabcolsep } {.1cm}
\tablewidth{0pt}
\tablecolumns{6}
\tablecaption{Hierarchical Bayesian modeling results for subsamples of warm Jupiters.}
\tablehead{
\colhead{Sample} & \colhead{Range} & \colhead{Bin Size} & \colhead{Hyperprior} & \colhead{$\alpha$} & \colhead{$\beta$}  
}
\startdata
    All Warm Jupiters & $\cdots$ & 200 & Truncated Gaussian &  1.00$^{+0.09}_{-0.08}$  & 2.79$^{+0.28}_{-0.26}$\\
    All Warm Jupiters & $\cdots$ & 200 & log-Uniform &  1.04$^{+0.09}_{-0.09}$ &  2.94$^{+0.32}_{-0.29}$\\
    Sub-Solar Metallicity & [Fe/H] $<$ 0 & 64 & Truncated Gaussian &  1.01$^{+0.15}_{-0.13}$ &  3.82$^{+0.58}_{-0.55}$\\
    Super-Solar Metallicity & [Fe/H] $\geq$ 0& 136&  Truncated Gaussian &  0.98$^{+0.11}_{-0.10}$ &  2.35$^{+0.29}_{-0.27}$ & \\
    Sub-Solar Metallicity & [Fe/H] $<$ 0 & 64 & log-Uniform &  1.39$^{+0.24}_{-0.22}$  & 6.12$^{+1.20}_{-1.09}$ \\
    Super-Solar Metallicity & [Fe/H] $\geq$ 0& 136& log-Uniform &  1.00$^{+0.12}_{-0.11}$ &  2.43$^{+0.31}_{-0.29}$ \\ 
    Sub-Solar Mass & $M_{\star}$ $<$ 1.15 $M_{\odot}$ & 102& Truncated Gaussian &  0.95$^{+0.12}_{-0.11}$  & 2.71$^{+0.37}_{-0.35}$\\
    Super-Solar Mass & $M_{\star}$ $\geq$ 1.15 $M_{\odot}$ & 98 & Truncated Gaussian &  1.01$^{+0.13}_{-0.12}$ &  2.65$^{+0.37}_{-0.34}$ \\
    Sub-Solar Mass & $M_{\star}$ $<$ 1.15 $M_{\odot}$ & 102 & log-Uniform &  1.01$^{+0.13}_{-0.12}$  & 2.96$^{+0.45}_{-0.41}$\\
    Super-Solar Mass & $M_{\star}$ $\geq$ 1.15 $M_{\odot}$ & 98 & log-Uniform &  1.06$^{+0.14}_{-0.13}$ &  2.89$^{+0.44}_{-0.41}$ \\
    Orbital Period & 10 d $<$ $\mathit{P}$ $<$ 100 d& 85 & Truncated Gaussian &  0.85$^{+0.11}_{-0.10}$  & 2.37$^{+0.36}_{-0.33}$\\
    Orbital Period & 100 d $\le$ $\mathit{P}$ $<$ 365 d &115 & Truncated Gaussian &  1.09$^{+0.13}_{-0.12}$ &  2.96$^{+0.37}_{-0.35}$ \\
    Orbital Period & 10 d $<$ $\mathit{P}$ $<$ 100 d & 85 & log-Uniform &  0.88$^{+0.13}_{-0.11}$  & 2.55$^{+0.43}_{-0.38}$\\
    Orbital Period & 100 d $\le$ $\mathit{P}$ $<$ 365 d &115  & log-Uniform &  1.17$^{+0.15}_{-0.13}$ &  3.28$^{+0.47}_{-0.42}$ \\
\enddata
\tablecomments{Beta distribution model posterior constraints from MCMC fitting for various samples considered in this study. Median values and 68$\%$ credible intervals are reported. We also test an underlying Rayleigh and Gaussian models to test for model dependency (see Appendix \ref{sec:Additional_Statistical_Tests}). \label{tab:Model_Parameters}}
\end{deluxetable*}

\section{Discussion}\label{sec:Discussion}

\subsection{Warm Jupiter Detectability}\label{sec:Warm_Jupiter_Observability}

The heterogeneity of instrument precision, host star jitter levels, host star brightness, relatively broad range of host masses, number of RV observations, method of discovery, and time baseline of each RV dataset mean that each star in this sample has a different (and difficult to reconstruct) completeness level in mass and orbital separation.  In principle, this can imprint biases on the eccentricity results if these planet signals are close to the detection sensitivity threshold. However, these biases would be mitigated if planet signals are much higher than the instrument or astrophysical noise levels.  A full treatment of survey completeness is outside the scope of this study, but here we further discuss some of the characteristics of giant planets and host stars in our sample as they relate to potential biases.

Most stars in this survey were part of programs targeting nearby, slowly rotating, bright stars, either on the main sequence or on the evolved subgiant and giant branches. This selection helps minimize rotational broadening of absorption lines and stellar jitter, making these stars well-suited for precision radial velocity surveys. The median host star $V$-band magnitude in this analysis is 7.2~mag with 80$\%$ of the sample brighter than $V$ = 10~mag.

A 1 $M_\mathrm{Jup}$ warm Jupiter on a circular orbit around a Sun-like star at 1 AU induces a semi-amplitude of 28.4 m s$^{-1}$. This amplitude scales with the orbital and physical parameters as 
   \begin{equation}
K = 28.4\ {\rm m\,s^{-1}} 
    \left( \frac{1\ {\rm yr}}{P} \right)^{1/3} 
    \left( \frac{M_p \sin i}{M_{\rm Jup}} \right) 
    \left( \frac{M_\odot}{M_\star} \right)^{2/3}
    \left( \frac{1}{\sqrt{1 - e^2}} \right),
\end{equation}
    where $M_\star$ is the mass of the host star, $e$ is the orbital eccentricity, and $P$ is the orbital period of the planet (\citealt{Cumming1999}). 89$\%$ (178/200) of the warm Jupiters in this sample induce semi-amplitudes on their host stars of \emph{K} $\geq$ 20 m s$^{-1}$. Only 11$\%$ (22/200) of the systems exhibit semi-amplitudes \emph{K} $<$ 20 m s$^{-1}$ and only 1$\%$ (2/200) have semi-amplitudes \emph{K} $<$ 10 m s$^{-1}$. The median error on the RV measurements used in this sample is $\sigma_{RV}$ $\approx$ 5 m s$^{-1}$. All of the planetary signals are well above a low signal-to-noise threshold (\emph{K} $\gg$ $\sigma_{\mathrm{RV}}$), both for higher- and lower-mass host stars in the sample, so we expect eccentricity-related RV discovery biases to be minimal.

\subsection{Stellar Parameter Uncertainties and Potential Biases}\label{sec:Mass_Metallicity_Biases}
The masses and metallicities that we adopt in this study are inferred using a broad range of techniques. To test whether a homogeneous and uniform approach might impact our results, we cross-matched the sample in this analysis with the stellar parameters derived from Gaia's RVS spectra (detailed in Section \ref{sec:Host_Stars}). We then re-analyze the underlying eccentricity distributions as a function of stellar host mass and metallicity using the uniformly compiled Gaia RVS parameters. Using the same HBM framework, we apply the solar metallicity cutoff ([Fe/H] = 0) described in Section \ref{sec:Host_Star_Metallicity} to the Gaia [M/H] metallicity measurements, resulting in a sample of 180 systems: 86 metal-poor and 94 metal-rich stars. Similarly, we adopt the same stellar mass cutoff of 1.15 $M_{\star}$ from Section \ref{sec:Host_Star_Mass} based on Gaia FLAME mass estimates, yielding 138 systems: 76 lower-mass and 62 higher-mass stars. The results are in excellent agreement with our original analysis when comparing the the underlying eccentricity distributions using the adopted Gaia DR3 stellar mass and metallicity values for the host stars. This reinforces our earlier conclusion: there is evidence for a difference when analyzing stellar host metallicity, but no significant difference is observed as a function of stellar host mass within this sample.

\subsection{Discerning the Role of Stellar Mass and Metallicity in Warm Jupiter Migration}\label{sec:Mass_Vs_Metallicity}

Within the framework of the core accretion theory (\citealt{Pollack1996}), stellar mass and metallicity (and, by extension, disk mass and metallicity) have long been recognized as important parameters that are expected to be associated with the enhanced formation of Jovian-mass planets (\citealt{Hayashi1981}; \citealt{Laughlin2004}; \citealt{IdaLin2005}; \citealt{KennedyKenyon2008}). The positive correlation between host star mass and the occurrence rate of giant planets has been observed over a wide range of masses (\citealt{Johnson2007}; \citealt{Bowler2010}). Similarly, the relationship between giant planet formation efficiency and host star metallicity has been well studied (\citealt{Santos2004}; \citealt{FischerValenti2005}).

\citet{Johnson2010} disentangled host star mass and metallicity as determining factors that enhance giant planet formation. They concluded that both mass and metallicity increase the giant planet fraction in their sample and derived an empirical relationship between giant planet frequency, stellar host mass, and metallicity. In our analysis, we use Equation 8 of \citet{Johnson2010} to further interpret the excess of highly eccentric orbits around metal rich stars and the lack of apparent differences in stellar mass (Figure \ref{fig:beta_distributions}),

\begin{equation}\label{eqn:mass_metallicity_equation}
  f(M_\star,{\rm [Fe/H]}) = 0.07 \pm 0.01
  \times \left(\frac{M_\star}{M_\odot}\right)^{1.0 \pm 0.3} 
  \times 10^{1.2 \pm 0.2 {\rm [Fe/H]}}, 
\end{equation}

\noindent where [Fe/H] is the stellar metallicity and $M_\star$ is the stellar mass. 

If warm Jupiter eccentricities are closely connected to the occurrence rates of giant planets, perhaps via increased chances of scattering events, then we would expect the way in which $M_{\star}$ and [Fe/H] depend on \emph{f} to be reflected in the underlying eccentricities of this warm Jupiter sample. For instance, if the dynamic range of metallicity in our sample is very large, but the range for mass is small, we might see signs of differences in the eccentricity distribution for metallicity but not for mass, or alternatively, for mass but not for metallicity. We want to know how the expected giant planet occurrence rate scales in a relative sense with mass and metallicity given the bins of mass and metallicity used in our analysis in Table \ref{tab:Model_Parameters}.  Ultimately, do the observed differences or similarities in eccentricity distributions follow expected trends with giant planet occurrence rate?

Here, we calculate the expected intrinsic frequency of giant planets as a function of host star mass for the divisions of mass and metallicity adopted in this study.  First we consider the marginalized mass bins from Section \ref{sec:Host_Star_Mass}.  Among the stars with $M_{\star}$ $<$ 1.15 $M_{\odot}$, the average and standard deviation of the mass is 0.96 $\pm$ 0.14 $M_{\odot}$. Within that same bin, the average metallicity is 0.06 $\pm$ 0.24 dex. For the higher-mass bin of $M_{\star}$ $\geq$ 1.15 $M_{\odot}$, the average and standard deviation of the mass is 1.47 $\pm$ 0.29 $M_{\odot}$. Within that bin, the metallicity is 0.05 $\pm$ 0.19 dex. In both scenarios, we Monte Carlo all masses and metallicities using Equation \ref{eqn:mass_metallicity_equation}, which yields expected planet fractions of 0.07$^{+0.04}_{-0.06}$ and 0.11$^{+0.05}_{-0.07}$, respectively.

Next, we compute the frequency of giant planets in our sample as a function of host star metallicity using the bins from Section \ref{sec:Host_Star_Metallicity}. For the subset of stars with [Fe/H] $<$ 0 dex, the average metallicity is -0.20 $\pm$ 0.16 dex. Within that same bin, the average and standard deviation of the mass is 1.25 $\pm$ 0.39 $M_{\odot}$. For the metal-rich bin of [Fe/H] $\geq$ 0, the average metallicity is 0.18 $\pm$ 0.11 dex. Within that bin, the average and standard deviation of the mass is 1.18 $\pm$ 0.30 $M_{\odot}$. These results yield planet fractions of 0.05$^{+0.02}_{-0.03}$ and 0.13$^{+0.05}_{-0.06}$, respectively.

\begin{figure*}
  \centering
    \begin{minipage}[b]{0.48\textwidth}
        \centering
        \includegraphics[width=\linewidth]{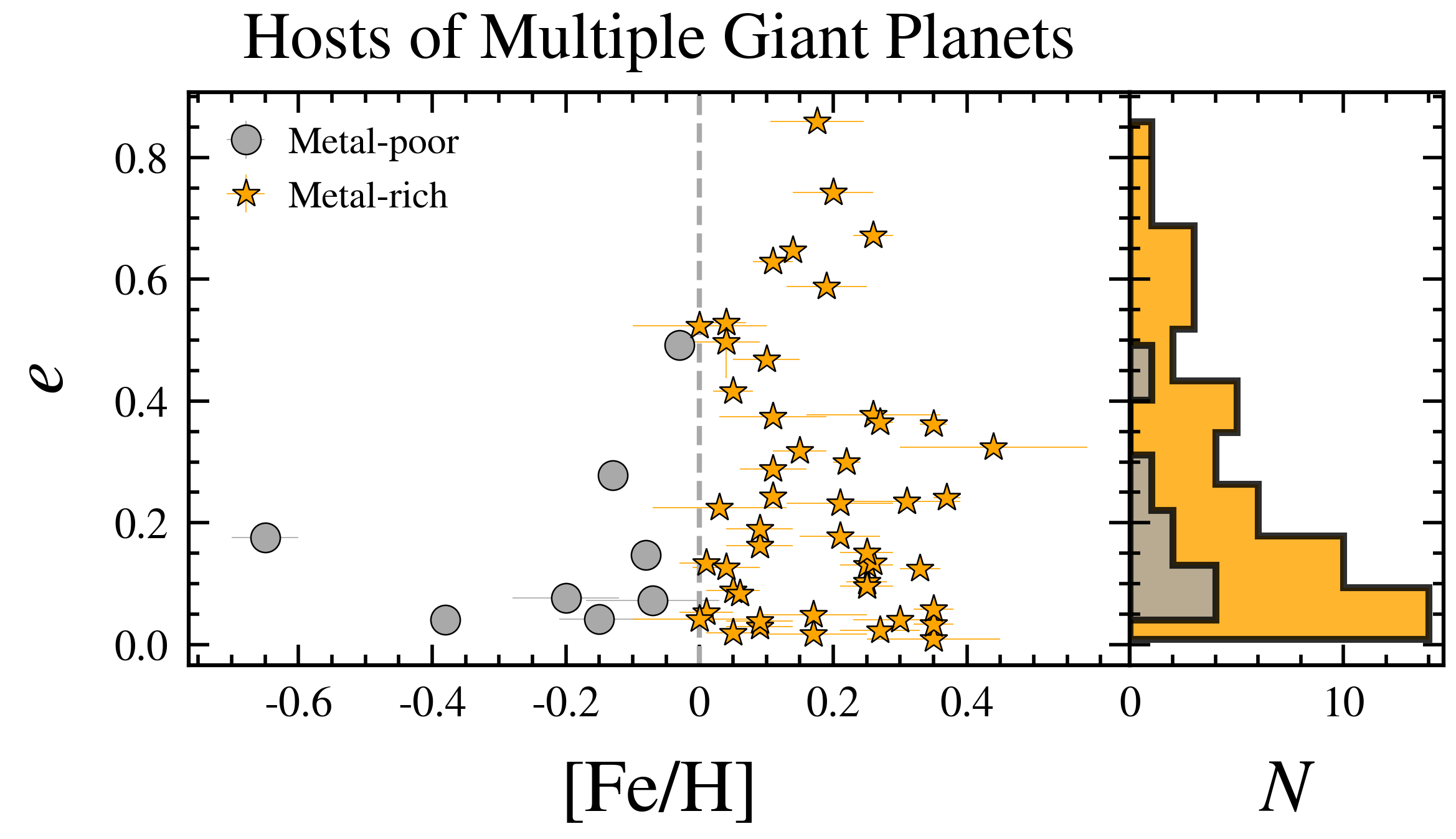}
    \end{minipage}
    \hfill
    \begin{minipage}[b]{0.48\textwidth}
        \centering
        \includegraphics[width=\linewidth]{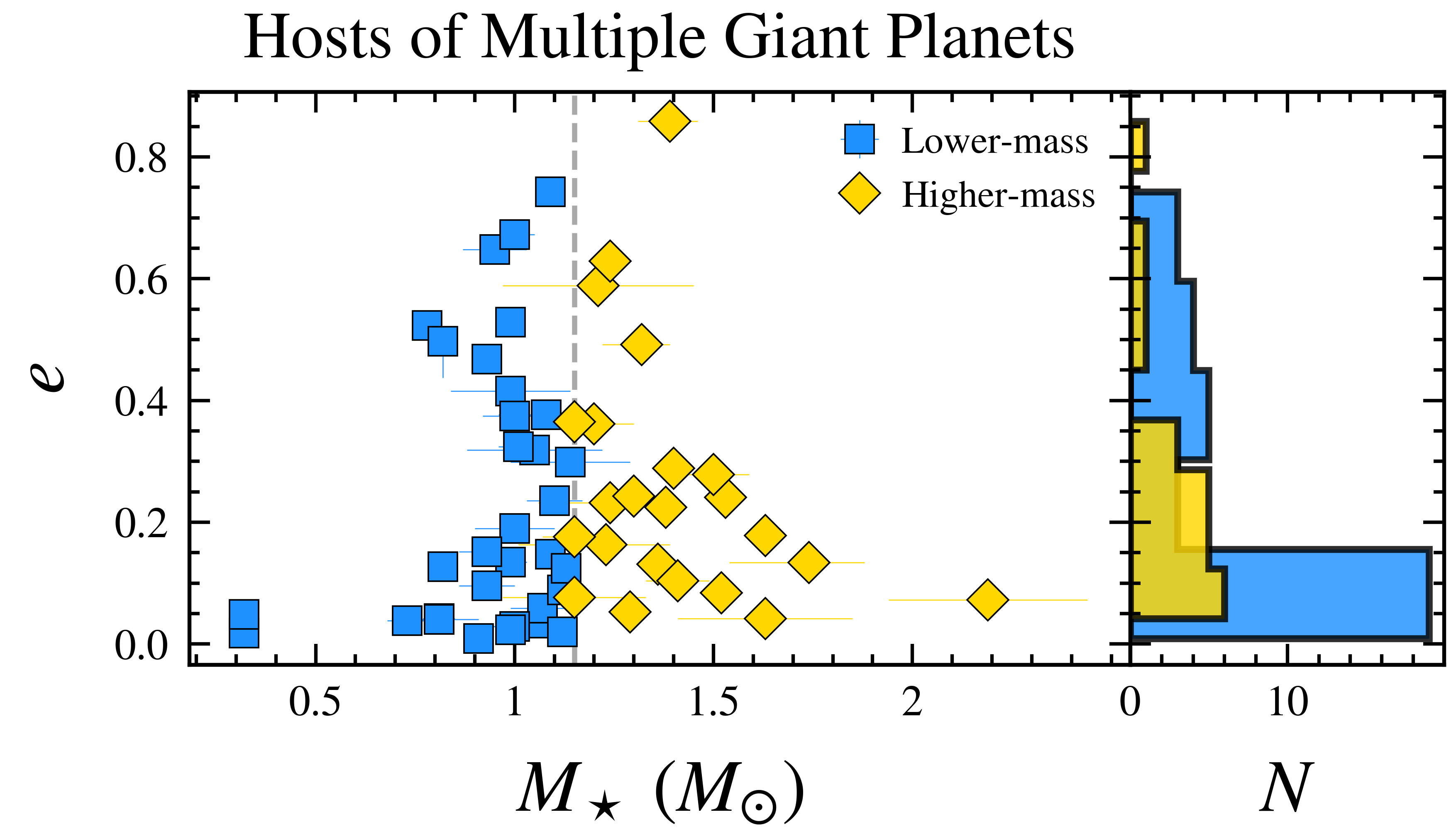}
    \end{minipage}
    \caption{Left: Stellar host metallicity plotted as a function of eccentricity for all warm Jupiters that reside in systems with at least one additional giant planet. Metal-poor ([Fe/H] $<$ 0) systems are in grey and metal-rich systems ([Fe/H] $\geq$ 0) are in orange. Right: Host star mass plotted as a function of eccentricity for all warm Jupiters that reside in systems with at least one additional giant planet. Lower-mass systems are in blue ($M_{\star}$ $<$ 1.15 $M_{\odot}$) and higher-mass systems are in gold ($M_{\star}$ $\geq$ 1.15 $M_{\odot}$) }
    \label{fig:multiple_giant_hosts}
\end{figure*}

Historically, the majority of Doppler surveys have focused on F, G, and K stars. The planets in our analysis have been collected from a diverse range of planet searches and surveys but reflect this focus on solar-type stars (see \citealt{Fischer2016} and \citealt{Perryman2018}), resulting in moderately restricted mass ranges of the host stars. Although the mass ranges are narrow, the metallicities within those mass bins are fairly broad. The two frequencies are marginally different at the 1.8-$\sigma$ level when analyzing the giant planet fraction as a function of host star metallicity. Conversely, when analyzing the giant planet fraction as a function of host star mass, the expected planet fractions are mutually consistent at the 0.6-$\sigma$ level. 

With a broader dynamic range in mass, or a larger sample size (which would result in greater precision in the inferred eccentricity distributions), this effect should imprint itself in planet eccentricities \emph{if} warm Jupiter eccentricities are mostly driven by planet scattering events, which themselves might scale with planet occurrence rate. For the stellar mass bins, we would expect the giant planet occurrence rate to increase by 57$\%$, and for the metallicity bins we would expect the occurrence rate to increase by 160$\%$.  

So within our sample, the impact of metallicity is expected to be a factor of $\sim$3 times more important. This could readily explain the significant difference in eccentricity distributions as a function of metallicity, and the lack of a distinction for stellar mass (the dynamic range of these two properties are unequal as seen in Figure \ref{fig:planet_properties}). Our results are therefore consistent with the dominant origin of warm Jupiter eccentricities being connected to giant planet occurrence rates, possibly through increased planet scattering events.  

A key assumption in this argument is that the giant planet multiplicity also scales in the same way as single giant planet frequency. That is, when stellar mass or metallicity increases, not only does the probability of finding a single giant planet increase, but the probability of having a second nearby planet susceptible to scattering also increases.  In short, this argument assumes that the probability distribution function for the presence of a second giant planet is independent of the first giant planet.

We also test the differences in the underlying eccentricity distributions in mass and metallicity using a Kolmogorov-Smirnov (KS) test, which can be used to assess whether two samples are consistent with originating from the same parent population. As the KS test does not traditionally include uncertainties, we use the MAP values of the individual eccentricity distributions as point estimates for this comparison. When assessing the underlying eccentricity distributions as a function of host metallicity we find a \textit{p}-value of 0.015. This tells that the null hypothesis---that the two samples are drawn from the same parent population---can be rejected as the $p$ value falls below the adopted threshold of 0.05.  This provides further support for a metallicity dependence of warm Jupiter eccentricity and is in line with our HBM results.

Next, we assess the underlying eccentricity distributions as a function of stellar host mass using a KS test. This yields a value of 0.55, and fails to reject the null hypothesis at a level more extreme than \textit{p}=0.05.  These tests give rise to similar conclusions as our HBM and KDE results that for this sample, warm Jupiter eccentricities do not seem to be correlated with stellar host mass.

\subsection{Multiple Giant Planets are More Common Around Metal-Rich Host Stars}\label{sec:planet_multiplicitiy_host_metallicity}

Using the The California Legacy Survey, \citet{Rosenthal2023} found that stellar hosts of multiple giant planets are more metal-rich than their single giant planet host counterparts. The implication is that the planet multiplicity fraction, not just the giant planet companion fraction, increases with stellar and protoplanetary disk metallicity. We find that among the population of 62 warm Jupiter hosts in our analysis that harbor more than one planet, 51 of these 62 stars host two or more giant planets with at least one being a warm Jupiter. 

Of these 51 multi-giant host stars, 8 are metal-poor ([Fe/H] $<$ 0) and 43 are metal-rich ([Fe/H] $\geq$ 0), further suggesting that metal-rich stars have a higher probability of producing more than one giant planet.\footnote{Note that stars in the sample have very different sensitivities to longer period planets and depends on time baseline of the RV observations, RV precision, and observing cadence. These numbers do not take into account sensitivity limits in the sample.} In other words, $P$($>$1 GP $|$ WJ, [Fe/H]$<$0.0 dex) = 17$^{+4}_{-6}\%$: about 17$\%$ of hosts harboring two or more giant planets, with at least one being a warm Jupiter, are metal-poor. Similarly, $P$($>$1 GP $|$ WJ, [Fe/H]$\geq$0.0 dex) = 83$^{+5}_{-5}\%$: about 83$\%$ of hosts harboring two or more giant planets, with at least one being a warm Jupiter, are metal-rich. As seen in Figure \ref{fig:multiple_giant_hosts}, stars hosting more than one giant planet in our sample are mostly metal-rich. One noticeable difference in both populations is the lack of highly eccentric planets in metal-poor systems hosting multiple giant planets (See Section \ref{sec:Host_Star_Metallicity}; Figure \ref{fig:beta_distributions}).

We also analyze the impact of stellar host mass on the eccentricities of warm Jupiters in systems that harbor more than one giant planet. Within our sample, we find no preference for high-mass ($M_{\star}$ $\geq$ 1.15 $M_{\odot}$) or low-mass ($M_{\star}$ $<$ 1.15 $M_{\odot}$) hosts to harbor more than one giant in systems where at least one of the giant planets is a warm Jupiter.
There are 28 warm Jupiters in this sample around lower-mass stars and 23 are around higher-mass stars (Figure \ref{fig:multiple_giant_hosts}). In both populations of systems hosting multiple giants, we find that there is no discernible difference in the eccentricities of the warm Jupiters, further suggesting that host star mass may not be impacting the underlying eccentricity distribution as significantly as host metallicity within 1 AU---at least for the ranges of stellar mass and metallicity considered in this study. (See Section \ref{sec:Host_Star_Mass}; Figure \ref{fig:beta_distributions}).


\section{Conclusion}\label{sec:Conclusion}
We present results of our uniform analysis of the underlying eccentricity distribution of warm Jupiters as a function of stellar mass, metallicity, and orbital separation.  There are several key takeaways:

\begin{itemize}[topsep=0pt, partopsep=0pt, itemsep=0pt, parsep=0pt]

\item From our uniform analysis of 200 warm Jupiters, we find that 27$^{+3}_{-4}\%$ of warm Jupiters detected through Doppler surveys have eccentricities consistent with \emph{near-circular} orbits (those with $e$ $<$ 0.1), and 73$^{+3}_{-3}\%$ of warm Jupiters have \emph{excited} eccentricities ($e$ $>$ 0.1).

\item Warm Jupiters take on a broad distribution of eccentricities. Using HBM, a Beta distribution describing the underlying eccentricity distribution of WJs produces $\alpha$ and $\beta$ values of 1.00$^{+0.09}_{-0.08}$ and 2.79$^{+0.28}_{-0.26}$. 

\item Among the sample of nearby Sun-like stars in this study, metallicity plays a larger role in shaping the underlying eccentricity distribution of warm Jupiters than stellar mass and final orbital distance---at least within 1 AU. 

\item Using empirically calibrated scaling relations between the occurrence rate of giant planets with stellar mass and metallicity, we find that for the ranges of stellar properties in this sample, we would expect the impact of metallicity to dominate over stellar mass by a factor of about 3.  Our results showing significant differences in the eccentricities of metal-rich and metal-poor stars, but no difference between higher- and lower-mass stars, is consistent with what would be expected if scattering is an important process shaping warm Jupiter eccentricities.

\item At the 3.3$\sigma$ level, warm Jupiters orbiting thick disk stars show signs of being less dynamically excited ($e$ $<$ 0.2) than warm Jupiters orbiting stars in the thin disk.

\item Altogether, the high prevalence of warm Jupiters with non-zero eccentricities and the way in which they depend on stellar mass and metallicity are consistent with planet scattering playing a dominant role in shaping the orbits of warm Jupiters, and perhaps delivering them to their current locations.

\end{itemize}

\section{acknowledgments}
We thank Kyle Franson, Lillian Jiang, and Danielle Berg for insightful conversations regarding the host stars and their stellar properties. We thank Gregory Gilbert, Erik Petigura, and Judah Van Zandt for valuable discussions regarding the likelihood modeling and the incorporation of the geometric transit bias.  B.P.B. acknowledges support from the National Science Foundation grant AST-1909209, NASA Exoplanet Research Program grant 20-XRP20$\_$2-0119, and the Alfred P. Sloan Foundation.
This research has made use of the NASA Exoplanet Archive, which is operated by the California Institute of Technology, under contract with the National Aeronautics and Space Administration under the Exoplanet Exploration Program and \citet{VizieR2000}, an online database with sources collected by the Centre de Données de Strasbourg (CDS).

\software{\texttt{numpy} \citep{Harris2020}, \texttt{matplotlib}  \citep{Hunter2007}, \texttt{ePop!} \citep{Nagpal2023}, \texttt{corner} \citep{Foreman-Mackey2016}, \texttt{emcee} \citep{Foreman-Mackey2013}, \texttt{ArviZ} \citep{arviz2019}, \texttt{radvel} \citep{Fulton2018}, \texttt{PyAstronomy} \citep{Czesla2019}, and \texttt{pandas} \citep{reback2020pandas}} 

\clearpage

\appendix
\section{Additional Statistical Tests}\label{sec:Additional_Statistical_Tests} 

\begin{figure}[h!]
 \hskip -0.8 in
 \gridline{\fig{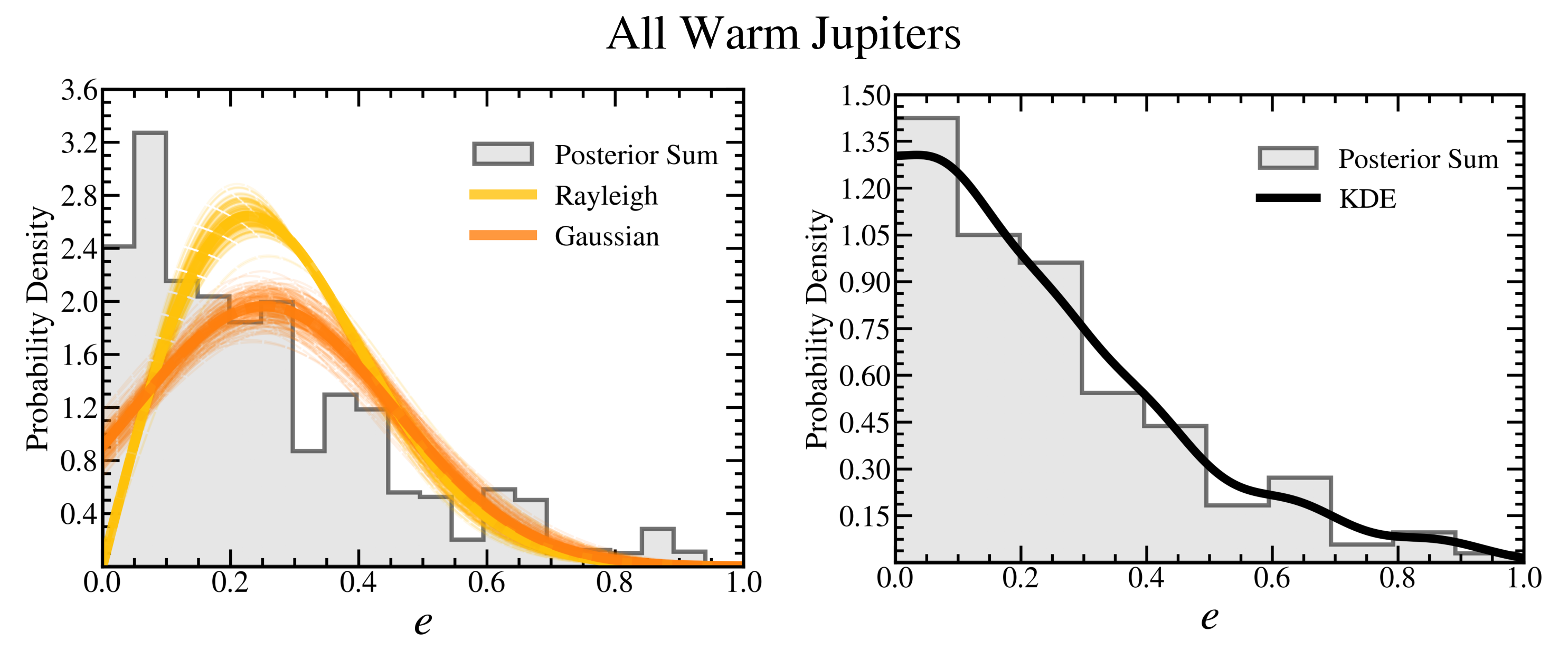}{0.9\textwidth}{}}
 \vskip -.3 in
 \gridline{\fig{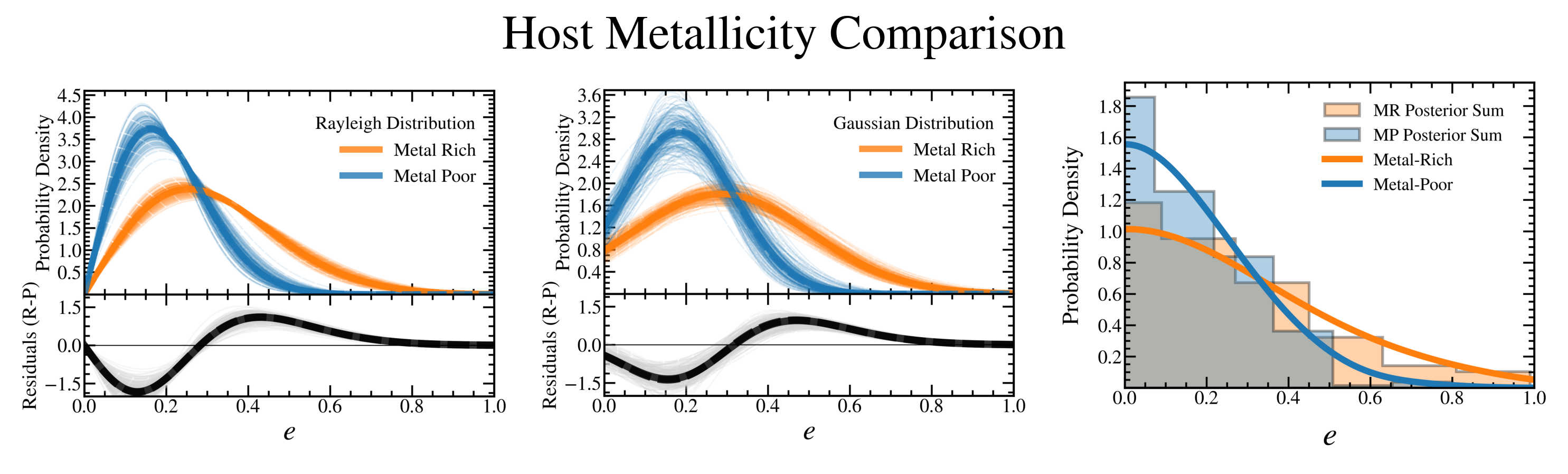}{0.9\textwidth}{}}
 \vskip -.3 in
\caption{\small Top Left: Additional underlying Rayleigh and Gaussian HBM tests for the full sample of warm Jupiters. Solid lines represent a Truncated Gaussian hyperprior while the dashed lines represent a log-uniform hyperprior. Thin lines in the background show randomly sampled distributions from the $\sigma_{R}$, $\mu_{G}$ and $\sigma_{G}$ hyperparameter posteriors. Top Right: KDE test for the full sample of warm Jupiters in our analysis. Bottom: Additional HBM tests for host star metallicity. We find that these additional tests are in good agreement with the differences between the underlying eccentricity distribution of warm Jupiters residing in metal-rich ([Fe/H] $\geq$ 0) and metal-poor ([Fe/H] $<$ 0) systems seen in Figure \ref{fig:beta_distributions}.
 \label{fig:additional_distributions1} }
\end{figure}

\begin{figure}
 \hskip -0.8 in
 \gridline{\fig{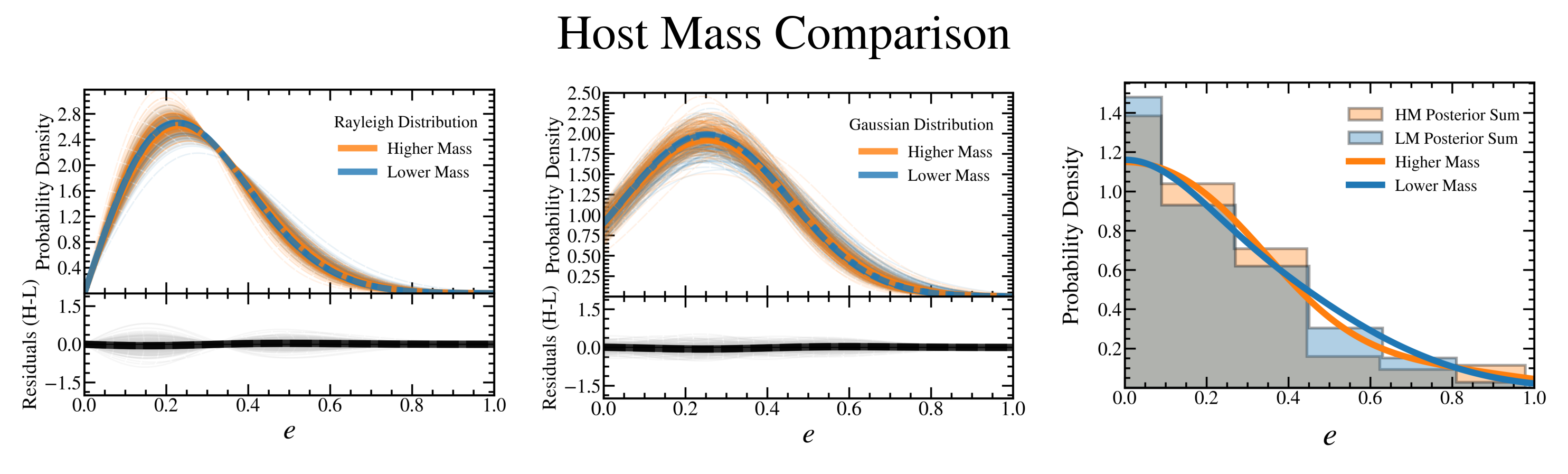}{0.9\textwidth}{}}
 \vskip -.3 in
 \gridline{\fig{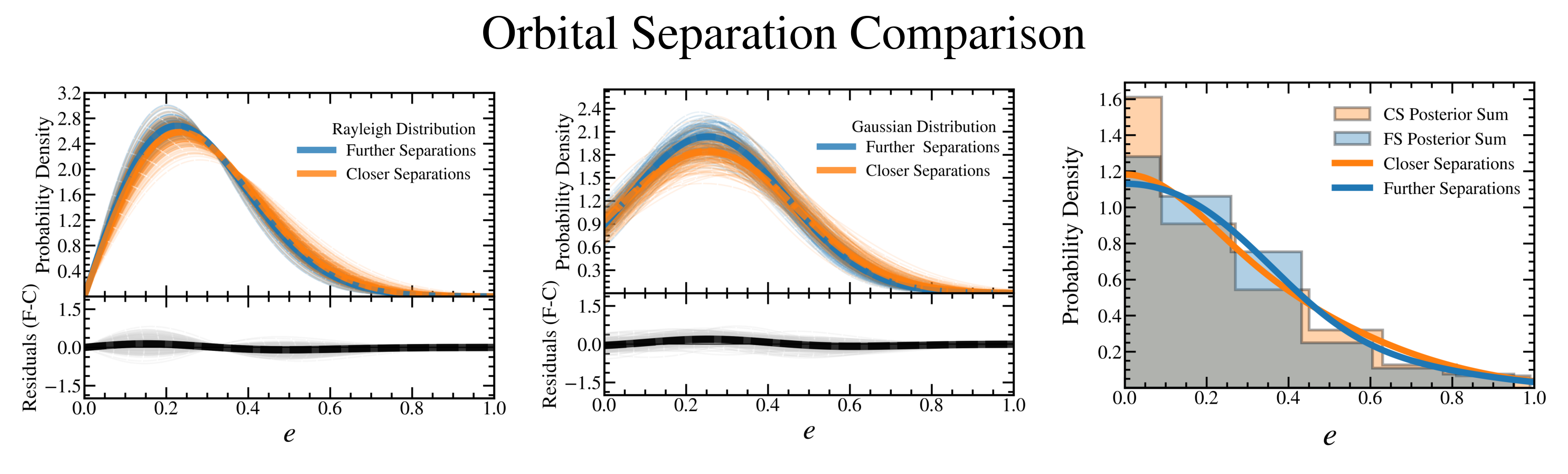}{0.9\textwidth}{}}
 \vskip -.3 in
\caption{\small Top: Additional tests for host star mass. These additional tests are in good agreement with the similarities between the underlying eccentricity distributions of warm Jupiters orbiting higher-mass host stars with $M_{\star}$ $\geq$ 1.15 $M_{\odot}$ and lower-mass host stars with $M_{\star}$ $<$ 1.15 $M_{\odot}$ seen in Figure \ref{fig:beta_distributions}. Bottom: Additional tests run for planets orbiting at farther separations (100 d $\le$ $\mathit{P}$ $<$ 365 d)  and closer separations (10 d $<$ $\mathit{P}$ $<$ 100 d) from their host stars. These tests reveal that the two underlying eccentricity distributions are in good agreement as seen in Figure \ref{fig:beta_distributions}. (See Figure \ref{fig:additional_distributions1} for additional details).
 \label{fig:additional_distributions2}}
\end{figure}

Here we report additional statistical tests to check for underlying model and hyperprior dependence when analyzing host star metallicity, mass, and orbital separation. To assess the impact of underlying models and hyperpriors, we run additional Hierarchical Bayesian tests with an underlying Rayleigh and Gaussian distribution. Both models are fit with a truncated Gaussian and log-Uniform hyperprior. In addition to the HBM, we estimate the underlying eccentricity distribution with a Gaussian Kernel density estimation (KDE). The KDE is calculated for 10$^{3}$ samples drawn from each eccentricity posterior distribution. Each KDE has an estimator bandwidth of 0.2 and is plotted alongside of the full eccentricity distribution as seen in Figures \ref{fig:additional_distributions1} and \ref{fig:additional_distributions2}. Tests are conducted on the full warm Jupiter sample and as a function of host star metallicity, mass, and orbital separation. 

We find that the underlying Rayleigh and Gaussian distributions and KDE's are in good agreement with the results of the Beta distribution HBM tests. Results of the two additional HBM tests can be found in Table \ref{tab:Appendix_Model_Parameters}. The Rayleigh distribution parameter is denoted by $\sigma_{R}$ and the Gaussian parameters are displayed as $\mu_{G}$ and $\sigma_{G}$. Median values and 68$\%$ credible intervals are reported.

\begin{deluxetable}{lccccccc}
\renewcommand\arraystretch{0.7}
\tabletypesize{\scriptsize}
\setlength{ \tabcolsep } {.1cm}
\tablewidth{0pt}
\tablecolumns{7}
\tablecaption{Raleigh and Gaussain distribution model posterior constraints from MCMC fitting for various samples considered in this study. Median values and 68$\%$ credible intervals are reported. \label{tab:Appendix_Model_Parameters} }
\tablehead{
\colhead{Sample} & \colhead{Range} & \colhead{Bin Size} & \colhead{Hyperprior} & \colhead{$\sigma_{R}$} & \colhead{$\mu_{G}$} & \colhead{$\sigma_{G}$}  
}
\startdata
    All Warm Jupiters & $\cdots$ & 200 & Truncated Gaussian & 0.23$^{+0.01}_{-0.01}$ & 0.25$^{+0.01}_{-0.01}$ &  0.20$^{+0.01}_{-0.01}$\\
    All Warm Jupiters & $\cdots$ & 200 & log-Uniform &  0.23$^{+0.01}_{-0.01}$ & 0.25$^{+0.01}_{-0.01}$ &  0.20$^{+0.01}_{-0.01}$\\
    Sub-Solar Metallicity & [Fe/H] $<$ 0 & 64 & Truncated Gaussian &  0.16$^{+0.01}_{-0.01}$ &  0.18$^{+0.02}_{-0.02}$ & 0.14$^{+0.01}_{-0.01}$\\
    Super-Solar Metallicity & [Fe/H] $\geq$ 0& 136&  Truncated Gaussian &  0.26$^{+0.01}_{-0.01}$ &  0.29$^{+0.02}_{-0.02}$ & 0.22$^{+0.01}_{-0.01}$\\
    Sub-Solar Metallicity & [Fe/H] $<$ 0 & 64 & log-Uniform &  0.16$^{+0.01}_{-0.01}$ &  0.18$^{+0.02}_{-0.02}$ & 0.13$^{+0.01}_{-0.01}$ \\
    Super-Solar Metallicity & [Fe/H] $\geq$ 0& 136& log-Uniform &  0.25$^{+0.01}_{-0.01}$ &  0.28$^{+0.02}_{-0.02}$ & 0.22$^{+0.01}_{-0.01}$ \\ 
    Sub-Solar Mass & $M_{\star}$ $<$ 1.15 $M_{\odot}$ & 102& Truncated Gaussian &  0.23$^{+0.01}_{-0.01}$ &  0.25$^{+0.02}_{-0.02}$ & 0.20$^{+0.02}_{-0.01}$\\
    Super-Solar Mass & $M_{\star}$ $\geq$ 1.15 $M_{\odot}$ & 98 & Truncated Gaussian &  0.23$^{+0.01}_{-0.01}$ &  0.25$^{+0.02}_{-0.02}$ & 0.21$^{+0.02}_{-0.01}$ \\
    Sub-Solar Mass & $M_{\star}$ $<$ 1.15 $M_{\odot}$ & 102 & log-Uniform &  0.23$^{+0.01}_{-0.01}$ &  0.25$^{+0.02}_{-0.02}$ & 0.20$^{+0.01}_{-0.01}$\\
    Super-Solar Mass & $M_{\star}$ $\geq$ 1.15 $M_{\odot}$ & 98 & log-Uniform &  0.23$^{+0.01}_{-0.01}$ &  0.25$^{+0.02}_{-0.02}$ & 0.21$^{+0.02}_{-0.01}$ \\
    Orbital Period & 10 d $<$ $\mathit{P}$ $<$ 100 d& 85 & Truncated Gaussian &  0.24$^{+0.01}_{-0.01}$ &  0.25$^{+0.02}_{-0.02}$ & 0.22$^{+0.02}_{-0.02}$\\
    Orbital Period & 100 d $\le$ $\mathit{P}$ $<$ 365 d &115 & Truncated Gaussian & 0.23$^{+0.01}_{-0.01}$ &  0.25$^{+0.02}_{-0.02}$ & 0.20$^{+0.01}_{-0.01}$ \\
    Orbital Period & 10 d $<$ $\mathit{P}$ $<$ 100 d & 85 & log-Uniform &  0.23$^{+0.01}_{-0.01}$ &  0.25$^{+0.02}_{-0.02}$ & 0.22$^{+0.02}_{-0.02}$\\
    Orbital Period & 100 d $\le$ $\mathit{P}$ $<$ 365 d&115  & log-Uniform &  0.23$^{+0.01}_{-0.01}$ &  0.25$^{+0.02}_{-0.02}$ & 0.20$^{+0.01}_{-0.01}$ \\
\enddata
\end{deluxetable}

\clearpage
\section{Joint Posterior Distributions}\label{sec:joint_posterior_distributions}

\begin{figure}[h!]
\begin{center}
{\includegraphics[width=\linewidth]{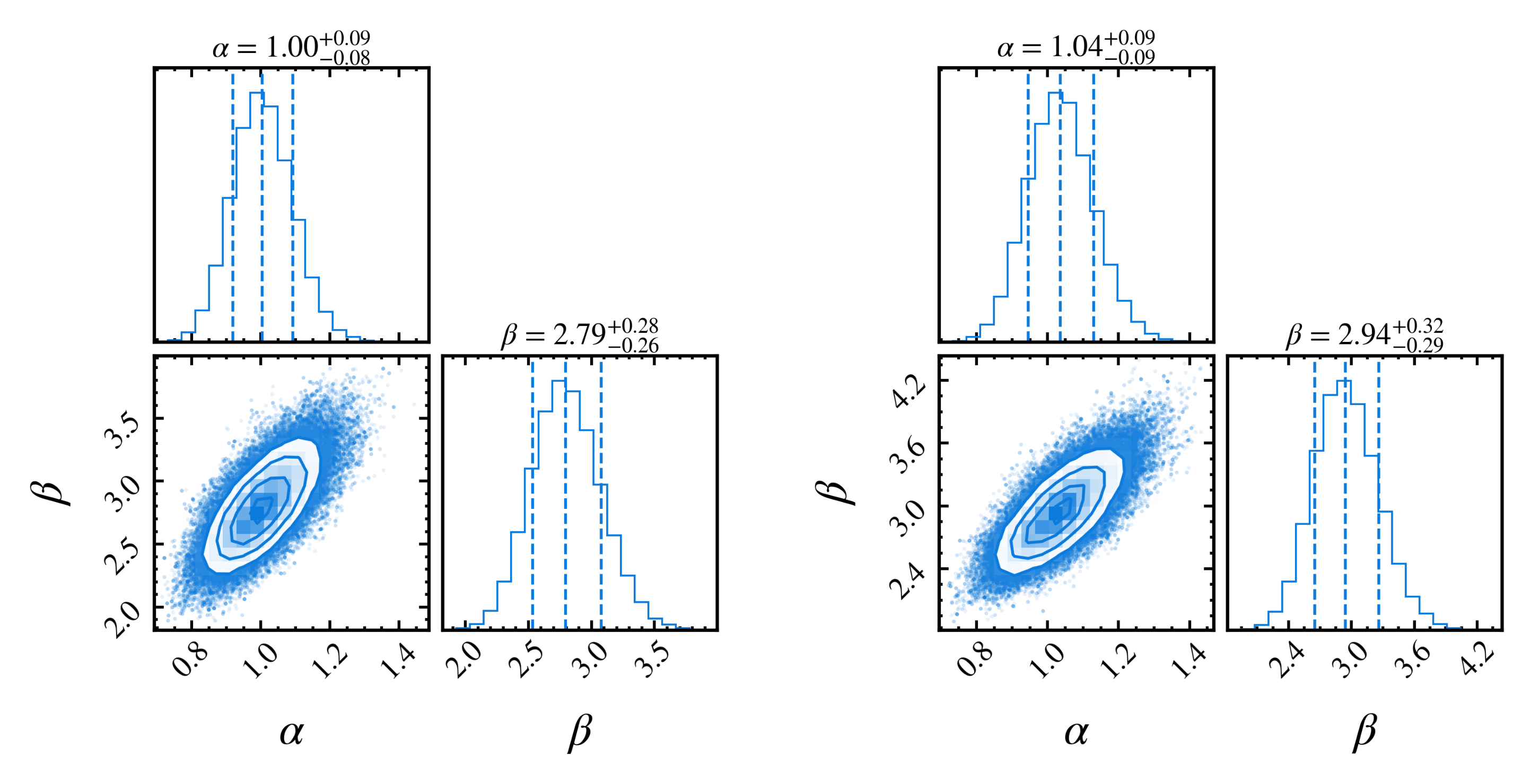}}
\caption{ Joint posterior distributions of the Beta distribution hyperparameters $\alpha$ and $\beta$ for the full sample of warm Jupiters. Left: The warm Jupiter underlying eccentricity distribution modeled with a Truncated Gaussian hyperprior. Right: The warm Jupiter underlying eccentricity distribution modeled with a log-uniform hyperprior.} 
\label{fig:corner_plots}
\end{center}
\end{figure}

Here we show the joint posterior distributions between the hyperparameters $\alpha$ and $\beta$ of our underlying Beta distribution. The results in Figure \ref{fig:corner_plots} illustrate the correlation between the $\alpha$ and $\beta$  hyperparamaters used to constrain the full warm Jupiter eccentricity distribution as displayed in Figure \ref{fig:True_WJ_Distribution}.

\bibliography{sample63}{}
\bibliographystyle{aasjournal}

\end{document}